\documentclass[vecphys,openany]{svmult}
\usepackage{graphicx}
\usepackage{amssymb}
\usepackage{cite}
\input epsf
\makeindex

\newcommand{\0}{\mbox{\bf 0}}

\newcommand{\s}{\mbox{\boldmath$\sigma$}}
\newcommand{\sk}{\s_{k}}
\newcommand{\kkk}{_{k}}
\newcommand{\bfe}{\mbox{\boldmath$e$}}
\newcommand{\bfg}{\mbox{\boldmath$g$}}
\newcommand{\bfk}{\mbox{\boldmath$k$}}
\newcommand{\bfp}{\mbox{\boldmath$p$}}
\newcommand{\bfq}{\mbox{\boldmath$q$}}
\newcommand{\bfu}{\mbox{\boldmath$u$}}
\newcommand{\bfx}{\mbox{\boldmath$x$}}
\newcommand{\bfA}{\mbox{\boldmath$A$}}
\newcommand{\bfF}{\mbox{\boldmath$F$}}
\newcommand{\bfN}{\mbox{\boldmath$N$}}
\newcommand{\ek}{\bfe_{k}}
\newcommand{\qk}{\bfq_{k}}
\newcommand{\uk}{\bfu_{k}}
\newcommand{\gk}{\bfg_{k}}

\begin{document}

\title*{Hydrodynamics from Grad's equations: What can we learn from exact solutions?}
\label{exactGrad}
\author{Iliya V. Karlin \and Alexander N. Gorban}
\titlerunning{Hydrodynamics from Grad's equations: exact solutions}

\institute{ETH Z\"urich, Switzerland \and University of Leicester, United Kingdom}
\date{}
\maketitle

\abstract{ A detailed treatment of the classical Chapman--Enskog derivation of
hydrodynamics is given in the framework of Grad's moment equations. Grad's systems are
considered as the minimal kinetic models where the Chapman--Enskog method can be studied
exactly, thereby providing the basis to compare various approximations in extending the
hydrodynamic description beyond the Navier--Stokes approximation. Various techniques,
such as the method of partial summation, Pad\'e approximants, and invariance principle
are compared both in linear and nonlinear situations.}

\section{The ``ultra-violet catastrophe" \\ of the Chapman--Enskog expansion} \label{UVC}

Most of the interesting expansions in non-equilibrium statistical
physics are divergent. This paraphrase of the well known folklore
\index{Dorfman's theorem}``Dorfman's theorem" conveys the
intrinsic problem of many-body systems: A number of systematic (at
the first glance) methods has led to
\begin{itemize}
\item{An excellent but already known on the phenomenological
grounds first approximation;} \item{Already the next correction,
not known phenomenologically and hence of interest, does not exist
because of divergence.}
\end{itemize}
There are many examples of this situations: Cluster expansion of
the exact collision integral for dense gases leads to divergent
approximations of transport coefficients, non-convergent long
tails of correlation functions in the Green--Kubo formulae etc.

The derivation of the hydrodynamic equations from a microscopic
description is the classical problem of physical kinetics. As is
well known, the famous Chapman--Enskog method \cite{Chapman}
provides an opportunity to compute a solution from the Boltzmann
kinetic equation as a formal series in powers of the Knudsen
number $\epsilon$.  The parameter $\epsilon$ reflects the ratio
between the mean free path of a particle, and the scale of
variations of the hydrodynamic fields (density, mean flux, and
temperature). If the Chapman--Enskog expansion is truncated at a
certain order, we obtain subsequently: the Euler hydrodynamics (
$\epsilon^0$), the Navier--Stokes hydrodynamics ($\epsilon^1$),
the Burnett hydrodynamics ($\epsilon^2$), the super-Burnett
hydrodynamics ($\epsilon^3$), etc. The post-Navier--Stokes terms
extend the hydrodynamic description beyond the strictly
hydrodynamic limit $\epsilon\ll 1$.

However, as it has been first demonstrated by \index{Bobylev's
instability}Bobylev \cite{Bob}, even in the simplest regime
(one-dimensional linear deviations around the global equilibrium),
the Burnett hydrodynamic equations violate the basic physics
behind the Boltzmann equation. Namely, sufficiently short acoustic
waves are amplified with time instead of decaying. This
contradicts the $H$-theorem, since all near-equilibrium
perturbations must decay. The situation does not improve in the
next, super-Burnett approximation.

This ``ultra-violet catastrophe" which occurs  in the lower-order
truncations of the Chapman--Enskog expansion creates therefore
very serious difficulties in the problem of an extension of the
hydrodynamic description into a highly non-equilibrium domain (see
\cite{Cercignani} for a discussion of other difficulties of the
post-Navier--Stokes terms of the Chapman--Enskog expansion). The
Euler and the Navier--Stokes approximations remain  basic in the
hydrodynamic description, while the problem of their extension is
one of the central open problems of kinetic theory. The study of
approximate solutions based on the Chapman--Enskog method still
continues \cite{Bob2}.

All this begs for a question: {\it What is wrong with the
Chapman--Enskog method?} At first glance, the failure of the
Burnett and of the super-Burnett hydrodynamics may be accounted in
favor of a frequently used argument about the asymptotic character
of the Chapman--Enskog expansion. However, it is worthwhile to
notice here that divergences in the low-order terms of formal
expansions are not too surprising. In many occasions, in
particular, in quantum field theory \cite{Jaffe} and in
statistical physics \cite{Parisi}, the situation is often improved
if one takes into account the very remote terms of the
corresponding expansions. Thus, a more constructive viewpoint on
the Chapman--Enskog expansion could be to proceed along these
lines, and to try to {\it sum up} the Chapman--Enskog series, at
least formally and approximately.

An attempt of this kind of working with the Chapman--Enskog
expansion is undertaken in this chapter. The formalities are known
to be  rather awkward for the Boltzmann equation, and untill now,
exact summations of the Chapman--Enskog expansion are known in a
very limited number of cases \cite{Hauge}. In this chapter, we
shall concentrate on the Chapman--Enskog method as applied to the
well known Grad moment equations \cite{Grad}.

The use of the Grad equations for our purpose brings, of course,
considerable technical simplifications as compared to the case of
the Boltzmann equation but it does not make the problem  trivial.
Indeed, the Chapman--Enskog method amounts to a  nonlinear
recurrence procedure even when applied to  the simplest,
linearized Grad equations. Moreover, as we shall see soon, the
Chapman--Enskog expansion for moment systems inherits Bobylev's
instability in the low-order approximations. Still, the advantage
of our approach is that many explicit results can be obtained and
analyzed. In order to summarize, in this chapter we consider
Grad's moment equations as finitely-coupled kinetic models where
the problem of reduced description is meaningful,  rather than as
models of extended hydrodynamics. The latter viewpoint is well
known as a microscopic background of the extended irreversible
thermodynamics \cite{Mueller,Jou}.

The outline of this chapter is as follows: after an introduction
of the Chapman--Enskog procedure for the linearized Grad equations
(Subsect. \ref{G_intro}), we shall start the discussion  with two
examples (the linearized one- and three-dimensional 10 moment Grad
equations) where the Chapman--Enskog series is summed up exactly
in closed form (sections \ref{exact_10_1} and \ref{exact_10_3}).
These results makes it possible to discuss the features of the
Chapman--Enskog solution in the short-wave domain in the framework
of the model, and  will serve the purpose of testing various
approximate methods thereafter. We shall see, in particular, that
the ``smallness" of the Knudsen number $\epsilon$ used to develop
the Chapman--Enskog method has no direct meaning in the exact
result. Also, it will become clear that finite-order truncations,
even provided they are stable, give less opportunities to
approximate the solution in a whole, and especially in the
short-wave domain.

The exact solutions are, of course, the lucky exceptions, and even
for the Grad moment equations the complexity of the
Chapman--Enskog method increases rapidly with an increase of the
number of the moments taken into account. Further (section
\ref{PS})  we shall review a technique of summing the
Chapman--Enskog expansion {\it partially}. This technique is
heuristic (as are the methods of partial summing in general), but
it still removes the Bobylev instability, as well as it
qualitatively reproduces the features of the exact solutions in
the short-wave limit.

The approach of working in the sections mentioned so far  falls
into the paradigm of the Taylor-like expansions into powers of the
Knudsen number. This viewpoint on the problem of the derivation of
the hydrodynamics will be {\it altered} beginning with Sect.
\ref{DI_intro}. There we demonstrate that a condition of a {\it
dynamic invariance} which can be realized directly and with no
restrictions of the Knudsen number brings us to the same result as
the exact summation of the Chapman--Enskog expansion. The
Chapman--Enskog method thereafter can be regarded as {\it one}
possibility to solve the resulting invariance equations. Further,
we demonstrate that iterative methods provide a reasonable
alternative to the Taylor expansion in this problem. Namely, we
show that the Newton method has certain advantages over the
Chapman--Enskog method (Sect. \ref{DI_N}). We also establish a
relationship between the method of partial summation  and the
Newton method.

The material of further sections serves for an illustrative
introduction how the pair ``invariance equation + Newton method"
can be applied to problems of kinetic theory. The remaining
sections of this chapter are devoted to further examples of this
approach on the level of the Grad equations. In sections
\ref{DI_13_1} and \ref{DI_13_3} we derive and discuss the
invariance equations for the linearized thirteen-moment Grad
equations. Section \ref{DIP} is devoted to kinetic equations of
the Grad type, arising in problems of phonon transport in massive
solids at low temperatures. In particular, we demonstrate that the
onset of the second sound regime of phonon propagation corresponds
to a branching point of the exact sum of the relevant
Chapman--Enskog expansion.

In Sect. \ref{NL} we apply the invariance principle to nonlinear
Grad equations. We sum up exactly a {\it subseries} of the
Chapman--Enskog expansion, namely, the dominant contribution in
the limit of high average velocities. This type of contribution is
therefore important for an extension of the hydrodynamic
description into the domain of strong shock waves. We present a
relevant analysis of the corresponding invariance equation, and,
in particular, discuss the nature of singular points of this
equation. A brief discussion concludes this chapter. Some of the
results presented below were published earlier in
\cite{GKJETP91,GKTTSP92,KTTSP92,book,GKPRL96,KDNPRE97,K2,JPA00},
and summarized in \cite{KGAnPh2002}.

\section[The Chapman--Enskog method \\ for linearized Grad's equations]
{The Chapman--Enskog method \\ for linearized Grad's equations}
\label{G_intro}
\sectionmark{The Chapman--Enskog method for linearized Grad's equations}

In this section, for the sake of completeness, we introduce
linearized Grad's equations and the Chapman--Enskog method for
them in the form that will used in the rest of this chapter. Since
the Chapman--Enskog method is extensively discussed in a number of
books, especially, in the classical monograph \cite{Chapman}, our
presentation will be brief.

The notation will follow that of the papers \cite{Bob,GKJETP91}.
We denote $\rho_0$, $T_0 $ and $\bfu=0$  the fixed equilibrium
values of density, temperature and averaged velocity (in the
appropriate Galilean reference frame), while $\delta\rho$, $\delta
T$ and $\delta{\bfu}$ are small deviations of the hydrodynamic
quantities from their equilibrium values. Grad's moment equations
\cite{Grad} which will appear below, contain the
temperature-dependent viscosity coefficient, $\mu(T)$. It is
convenient to write $\mu(T)=\eta(T)T$. The functional form of
$\eta(T)$ is dictated by the choice of the model for particle
interaction. In particular, we have $\eta={\rm const}$ for
Maxwell's molecules, and $\eta\sim \sqrt{T}$ for hard spheres.

We use the system of units in which Boltzmann's constant $k_{\rm
B}$ and the particle mass $m$ are equal to one. Let us introduce
the following system of dimensionless variables:
\begin{eqnarray}
\label{AnnPhysvar} {\bfu}=\frac{\delta{\bfu}}{\sqrt{T_0}},\ \rho=\frac{\delta\rho}{\rho_0},\
T=\frac{\delta T}{T_0},\\\nonumber \bfx=\frac{\rho_0}{\eta(T_0)\sqrt{T_0}}\bfx^{\prime}, \
t=\frac{\rho_0}{\eta(T_0 )}t^{\prime},
\end{eqnarray}
where ${\bfx}^{\prime}$ are spatial coordinates, and $t^{\prime}$
is time. Three-dimensional thirteen moment Grad's equations,
linearized near the equilibrium, take the following form when
written in terms of the dimensionless variables
(\ref{AnnPhysvar}):
\begin{eqnarray}
\label{balanceequations}
\partial_t \rho &=&-\nabla\cdot{\bfu},\\\nonumber
\partial_t {\bfu} &=&-\nabla \rho-\nabla T -\nabla\cdot
\s ,\\\nonumber
\partial_t T &=&-\frac{2}{3}(\nabla\cdot{\bfu}+\nabla\cdot{\bfq})
,\\
\label{Grad133}
\partial_t \s &=&-\overline{\nabla {\bfu}}-\frac{2}{5}\overline{\nabla {\bfq}}-\s
,\\\nonumber
\partial_t  {\bfq}&=&-\frac{5}{2}\nabla T-\nabla\cdot\s -\frac{2}{3} {\bfq}.
\end{eqnarray}
In these equations, $\s({\bfx},t)$ and ${\bfq({\bfx},t)}$ are dimensionless
quantities corresponding to the stress tensor and to the heat
flux, respectively.
Further, the gradient $\nabla$ stands for the vector of spatial derivatives
$\partial/\partial{\bfx}$. The dot denotes the standard scalar product, while the
overline stands for a symmetric traceless dyad. In particular,
\[
\overline {{\nabla}{\bfu}}={\nabla}{\bfu}+({\nabla}{\bfu})^T
-\frac{2}{3}{I}{\nabla}\cdot{\bfu},\]
where ${I}$ is unit matrix.

Grad's equations (\ref{balanceequations}) and (\ref{Grad133}) is
the simplest model of a coupling of the hydrodynamic variables,
$\rho({\bfx},t)$, $T({\bfx},t)$ and ${\bfu}(\bfx,t)$, to the
non-hydrodynamic variables $\s ({\bfx},t)$ and ${\bfq}({\bfx},t)$.
The  problem of reduced description is to close the first three
equations (\ref{balanceequations}), and to get an autonomous
system for the hydrodynamic variables alone. In other words, the
non-hydrodynamic variables $\s ({\bfx},t)$ and ${\bfq}({\bfx},t)$
should be expressed in terms of $\rho({\bfx},t)$, $T({\bfx},t)$
and ${\bfu}(\bfx,t)$. The Chapman--Enskog method, as applied for
this purpose to Grad's system (\ref{balanceequations}) and
(\ref{Grad133}), involves the following steps:

First, we introduce a formal parameter $\epsilon$,
and write instead of equations (\ref{Grad133}):
\begin{eqnarray}
\label{Grad133e}
\partial_t \s &=&-\overline{\nabla {\bfu}}-\frac{2}{5}
\overline{\nabla {\bfq}}-\frac{1}
{\epsilon}\s,\\\nonumber
\partial_t{\bfq}&=&-\frac{5}{2}\nabla T-\nabla\cdot\s -\frac{2}{3\epsilon}{\bfq}.
\end{eqnarray}
Second, the Chapman--Enskog solution is found as a formal
expansions of the stress tensor and of the heat flux vector:
\begin{eqnarray}
\label{expansion}
{\s}&=&\sum_{n=0}^{\infty}\epsilon^{n+1}{\s}^{(n)};\\\nonumber
{\bfq}&=&\sum_{n=0}^{\infty}\epsilon^{n+1}{\bfq}^{(n)}.
\end{eqnarray}
The zero-order coefficients, ${\s}^{(0)}$ and ${\bfq}^{(0)}$, are:
\begin{equation}
\label{ns133}
{\s}^{(0)}=-\overline{\nabla {\bfu}},\quad {\bfq}^{(0)}=-\frac{15}{4}\nabla T.
\end{equation}
Coefficients of order $n\ge 1$ are found from the recurrence
procedure:
\begin{eqnarray}
\label{procedure133}
{\s}^{(n)}&=&-\left\{
\sum_{m=0}^{n-1}\partial_t^{(m)}\s^{(n-1-m)}+\frac{2}{5}\overline{\nabla {\bfq}^{(n-1)}}
\right\},
\\\nonumber
{\bfq}^{(n)}&=&-\frac{3}{2}\left\{\sum_{m=0}^{n-1}\partial_t^{(m)}{\bfq}^{(n-1-m)}
+\nabla\cdot\s^{(n-1)}\right\},
\end{eqnarray}
where $\partial_t^{(m)}$ are recurrently defined {\it
Chapman--Enskog operators}. They act on functions
$\rho({\bfx},t)$, $T({\bfx},t)$ and ${\bfu}(\bfx,t)$, and on their
spatial derivatives, according to the following rule:
\begin{eqnarray}
\label{operators133}
\partial_t^{(m)}D\rho&=&\left\{
\begin{array}{lcl}
-D\nabla\cdot{\bfu}&\quad\quad&m=0\\
0&\quad&m\ge 1\end{array}\right.;\\\nonumber
\partial_t^{(m)}DT&=&\left\{
\begin{array}{lcl}
-\frac{2}{3}D\nabla\cdot{\bfu}&\quad&m=0\\
-\frac{2}{3}D\nabla\cdot{\bfq}^{(m-1)}&\quad&m\ge 1\end{array}\right.;\\\nonumber
\partial_t^{(m)}D{\bfu}&=&\left\{
\begin{array}{lcl}
-D\nabla(\rho +T)&\quad&m=0\\
-D\nabla\cdot\s^{(m-1)}&\quad&m\ge 1\end{array}\right..
\end{eqnarray}
Here $D$ is an arbitrary differential operator with constant coefficients.

Given the initial condition (\ref{ns133}), the Chapman--Enskog
equations (\ref{procedure133}) and (\ref{operators133}) are
recurrently solvable. Finally, by terminating the computation at
the order $N\ge0$, we obtain the $N$th order approximations to the
expansions (\ref{expansion}), $\s_{N}$ and ${\bfq}_N $:
\begin{equation}
\label{truncation133}
{\s}_N =\sum_{n=0}^{N}\epsilon^{n+1}{\s}^{(n)},\quad
{\bfq}_N =\sum_{n=0}^{N}\epsilon^{n+1}{\bfq}^{(n)}.
\end{equation}
Substituting these expressions instead of the functions $\s$ and
${\bfq}$ in  (\ref{balanceequations}), we close the latter to give
the hydrodynamic equations of the order $N$. In particular, $N=0$
results in the Navier--Stokes approximation, $N=1$ and $N=2$ give
the Burnett and the super-Burnett approximations, respectively,
and so on.

Though the ``microscopic" features of Grad's moment equations are,
of course, much simpler in comparison to the Boltzmann equation,
the Chapman--Enskog procedure just described is not trivial. Our
purpose is to study explicitly the features of the gradient
expansions like (\ref{expansion}) in the highly non-equilibrium
domain, and, in particular, to find out to what extend the
finite-order truncations (\ref{truncation133}) approximate the
solution, and what kind of alternative strategies to find
approximations are possible. In the following, when referring to
Grad's equations, we use the notation $mDnM$, where $m$ is the
spatial dimension of the corresponding fields, and $n$ is the
number of these fields. For example, the above system is the
$3D13M$ Grad's system.

\section[Exact summation of the Chapman--Enskog expansion]
{Exact summation of the Chapman--Enskog expansion} \label{exact_10}
\sectionmark{Exact summation of the Chapman--Enskog expansion}
\subsection{The $1D10M$ Grad equations} \label{exact_10_1}

In this section, we start the discussion with the exact summation
of the Chapman--Enskog series for the simplest Grad's system, the
one-dimensional linearized ten-moment equations. Throughout the
section we use the hydrodynamic variables
$p(x,t)=\rho(x,t)+T(x,t)$ and $u(x,t)$, representing the
dimensionless deviations of the pressure and of the average
velocity from their equilibrium values (see (\ref{AnnPhysvar})).
The starting point is the linearized Grad's equations for $p$,
$u$, and $\sigma$, where $\sigma$ is the dimensionless
$xx$-component of the stress tensor:

\begin{eqnarray}
\label{Grad101}
\partial_t p &=&-\frac{5}{3}\partial_x u,\\\nonumber
\partial_t u &=&-\partial_x p -\partial_x \sigma,\\\nonumber
\partial_t \sigma &=&-\frac{4}{3}\partial_x u
-\frac{1}{\epsilon}\sigma.
\end{eqnarray}

The system of equations for three functions is derived from the
ten-moment Grad's system (see (\ref{Grad103}) below). Equations
(\ref{Grad101}) provides the simplest model of a coupling of the
hydrodynamic variables, $u$ and $p$, to the single
non-hydrodynamic variable $\sigma$, and corresponds to a heat
non-conductive case.

Our goal here is to  reduce the description, and  to get a closed
set of equations with respect to variables $p$ and $u$ only. That
is, we have to express the function $\sigma$ in the terms of
spatial derivatives of  $p$ and $u$.  The Chapman--Enskog method,
as applied to (\ref{Grad101}) results in the following series
representation:
\begin{equation}
\label{series101}
\sigma=\sum_{n=0}^{\infty}\epsilon^{n+1}\sigma^{(n)}.
\end{equation}
The coefficients $\sigma^{(n)}$ are obtained from the following
recurrence procedure \cite{GKJETP91}:
\begin{equation}
\label{procedure101}
\sigma^{(n)}=-\sum_{m=0}^{n-1}\partial_t^{(m)}\sigma^{(n-1-m)},
\end{equation}
where the Chapman--Enskog operators $\partial_t^{(m)}$ act on $p$,
$u$, and their spatial derivatives as follows:
\begin{eqnarray}
\label{operators101}
\partial_t^{(m)}\partial_x^l u&=&\left\{
\begin{array}{l@{\quad}l}
-\partial_x ^{l+1}p, & m=0\\
-\partial_x^{l+1} \sigma^{(m-1)}, & m\ge 1\end{array}\right.,
\\\nonumber
\partial_t^{(m)}\partial_x^{l}p&=&\left\{
\begin{array}{l@{\quad}l}
-\frac{5}{3}\partial_x^{l+1} u, & m=0\\
0, & m\ge 1\end{array}\right..
\end{eqnarray}
Here $l\ge 0$ is an arbitrary integer, and $\partial_x^0 =1$.
Finally,
\begin{equation}\label{ns101}
\sigma^{(0)}=-\frac{4}{3}\partial_x u,\end{equation}
which leads to
the Navier--Stokes
approximation of the stress tensor: $\sigma_{\rm NS}=\epsilon\sigma^{(0)}$.

Because of the somewhat involved structure of the recurrence
procedure (\ref{procedure101}) and (\ref{operators101}), the
Chapman--Enskog method is a nonlinear operation even in the
simplest model (\ref{Grad101}). Moreover, the Bobylev instability
is again present.

Indeed, computing the coefficients $\sigma^{(1)}$ and
$\sigma^{(2)}$ on the basis of (\ref{procedure101}), we obtain:
\begin{equation}
\label{burnett101}
\sigma_{\rm B}=\epsilon\sigma^{(0)}+\epsilon^2 \sigma^{(1)}=
-\frac{4}{3}(\epsilon\partial_x u +\epsilon^2 \partial^2_x p),
\end{equation}
and
\begin{equation}
\label{sburnett101}
\sigma_{\rm SB}=\epsilon\sigma^{(0)}+\epsilon^2 \sigma^{(1)}+\epsilon^3
\sigma^{(2)}=
-\frac{4}{3}(\epsilon\partial_x u +\epsilon^2 \partial^2_x p
+\frac{1}{3}\epsilon^3 \partial^3_x u),
\end{equation}
for the Burnett and the super-Burnett approximations,
respectively. Now we can substitute each of the approximations,
$\sigma_{\rm NS}$, $\sigma_{\rm B}$, and $\sigma_{\rm SB}$ for
$\sigma$ in the second equation of the set (\ref{Grad101}). The
equations thus obtained, together with the equation for density
$\rho$, form the closed systems of the hydrodynamic equations of
the Navier--Stokes, Burnett, and super-Burnett levels. To see the
properties of the resulting equations, we compute the dispersion
relation for the hydrodynamic modes. Using a new space-time scale,
$x^{\prime}=\epsilon^{-1} x$, and $t^{\prime}=\epsilon^{-1} t$,
and representing $u=u_k \varphi(x^{\prime},t^{\prime})$, and
$p=p_k \varphi(x^{\prime},t^{\prime})$, where
$\varphi(x^{\prime},t^{\prime})=\exp(\omega
t^{\prime}+ikx^{\prime})$, and $k$ is a real-valued wave vector,
we obtain the following dispersion relations $\omega(k)$ from the
condition of a non-trivial solvability of the corresponding linear
system with respect to $u_k$ and $p_k $:
\begin{equation}
\label{dispersionns101}
\omega_{\pm}=-\frac{2}{3}k^2 \pm \frac{1}{3}i|k|\sqrt{4k^2 -15},
\end{equation}
for the Navier--Stokes approximation,

\begin{equation}
\label{dispersionburnett101}
\omega_{\pm}=-\frac{2}{3}k^2 \pm \frac{1}{3}i|k|\sqrt{8k^2 +15},
\end{equation}
for the Burnett approximation (\ref{burnett101}), and
\begin{equation}
\label{dispersionsburnett101}
\omega_{\pm}=\frac{2}{9}k^2 (k^2 -3) \pm \frac{1}{9}i|k|\sqrt{4k^6 -
24k^4 -72k^2 -135},
\end{equation}
for the super-Burnett approximation (\ref{sburnett101}).

These examples demonstrate that the real part ${\rm Re}(
\omega_{\pm}(k))\le 0$ for the Navier--Stokes
(\ref{dispersionns101}) and for the Burnett
(\ref{dispersionburnett101}) approximations, for all wave vectors.
Thus, these approximations describe attenuating acoustic waves.
However, for the super-Burnett approximation, the function ${\rm
Re}( \omega_{\pm}(k))$ (\ref{dispersionsburnett101}) becomes
positive as soon as $|k|>\sqrt{3}$. That is, the equilibrium point
is stable within the Navier--Stokes and the Burnett approximation,
and it becomes  unstable within the super-Burnett approximation
for sufficiently short waves. Similar to the case of the Bobylev
instability of the Burnett hydrodynamics for the Boltzmann
equation, the latter result contradicts the dissipative properties
of the Grad system (\ref{Grad101}): the spectrum of the full
$1D10M$ system (\ref{Grad101}) is stable for arbitrary $k$.

Our goal now is to  sum up the series (\ref{series101}) in closed
form. Firstly, we should make some preparations.

As  demonstrated in \cite{GKJETP91} (see also below), the
functions $\sigma^{(n)}$ in (\ref{series101}) and
(\ref{procedure101}) have the following explicit structure to
arbitrary order $n\ge 0$:
\begin{eqnarray}
\label{structure101}
\sigma^{(2n)}&=&a_n \partial_x^{2n+1}u, \\\nonumber
\sigma^{(2n+1)}&=&b_n \partial_x^{2(n+1)}p,
\end{eqnarray}
where the coefficients $a_n$ and $b_n$ are determined through  the
recurrence procedure (\ref{procedure101}), and
(\ref{operators101}). The Chapman--Enskog procedure
(\ref{procedure101}) and (\ref{operators101}) can be represented
in terms of the real-valued coefficients $a_n$ and $b_n$
(\ref{structure101}).

Knowing the structure (\ref{structure101}) of the coefficients of
the Chapman--Enskog expansion (\ref{series101}), we can write down
its formal sum. It is convenient to use the Fourier variables
introduced above which amounts essentially to the change
$\epsilon\partial_x \rightarrow ik$. Substituting expression
(\ref{structure101}) into the Chapman--Enskog series
(\ref{series101}), we obtain the following formal expression for
the Fourier image of the sum:
\begin{equation}
\label{sigma101}
\sigma_{k}=ikA(k^2 )u_k -k^2 B(k^2 )p_k ,
\end{equation}
where the functions  $A(k^2)$ and $B(k^2)$ are formal power series
with the coefficients (\ref{structure101}):
\begin{eqnarray}
\label{AB101}
A(k^2 )&=&\sum_{n=0}^{\infty}a_n (-k^2 )^n , \\\nonumber
B(k^2 )&=&\sum_{n=0}^{\infty}b_n (-k^2 )^n.
\end{eqnarray}

Thus, the question of the summation of the Chapman--Enskog series
(\ref{series101}) amounts to finding the two functions, $A(k^2 )$
and $B(k^2 )$ (\ref{AB101}). Knowing them, the dispersion relation
for the hydrodynamic modes can be derived:
\begin{equation}
\label{dispersion101}
\omega_{\pm}=\frac{k^2 A}{2} \pm
\frac{|k|}{2}\sqrt{k^2 A^2 -\frac{20}{3} (1-k^2 B)}.
\end{equation}

We shall concentrate now on the problem of deriving $A(k^2 )$ and
$B(k^2 )$ (\ref{AB101}) in closed form. For this purpose, we shall
first express the Chapman--Enskog procedure (\ref{procedure101})
and (\ref{operators101}) in terms of the coefficients $a_n $ and
$b_n $ (\ref{structure101}). At the same time, our derivation will
constitute proof for the structure (\ref{structure101}).

It is convenient to start with the Fourier representation of
(\ref{procedure101}) and (\ref{operators101}). Writing $u=u_k
\exp(ikx)$, $p=p_k \exp(ikx)$, and $\sigma=\sigma_k \exp(ikx)$, we
obtain:
\begin{eqnarray}
\label{Foperators101}
\partial_t^{(m)} u_k &=&\left\{
\begin{array}{l@{\quad}l}
-ikp_k , & m=0\\
-ik \sigma_k^{(m-1)}, & m\ge 1\end{array}\right.,
\\\nonumber
\partial_t^{(m)}p_k &=&\left\{
\begin{array}{l@{\quad}l}
-\frac{5}{3}ik u_k , & m=0\\
0, & m\ge 1\end{array}\right.,
\end{eqnarray}
while
\begin{equation}
\label{Fprocedure101}
\sigma^{(n)}_k =-\sum_{m=0}^{n-1}\partial_t^{(m)}\sigma_k^{(n-1-m)},
\end{equation}
and

\begin{eqnarray}
\label{Fstructure101}
\sigma_k^{(2n)}&=&a_n (-k^2 )^n iku_k , \\\nonumber
\sigma_k^{(2n+1)}&=&b_n (-k^2 )^{n} (-k^2 )p_k .
\end{eqnarray}
The Navier--Stokes and the Burnett approximations give $a_0
=-\frac{4}{3}$, and $b_0 =-\frac{4}{3}$. Thus, the structure
(\ref{Fstructure101}) is proved for $n=0$.

The further derivation relies on  induction. Let us assume that
the ansatz (\ref{Fstructure101}) is proven up to the order $n$.
Computing the coefficient $\sigma_k^{(2(n+1))}$ from
(\ref{Fprocedure101}), we have:
\begin{equation}
\label{Fodd101}
\sigma_k^{(2(n+1))}=-\partial_t^{(0)}\sigma_k^{(2n+1)}-
\sum_{{m=0}}^{{n}}\partial_t^{(2m+1)}\sigma_k^{(2(n-m))}
-\sum_{{m=1}}^{{n}}\partial_t^{(2m)}\sigma_k^{(2(n-m)+1)}.
\end{equation}
Due to the assumption of the induction, we can adopt the form of
the coefficients $\sigma_k^{(j)}$ (\ref{Fstructure101}) in all the
terms on the right hand side of (\ref{Fodd101}). On the basis of
(\ref{Fstructure101}) and (\ref{Foperators101}), we conclude that
each term in the last sum of (\ref{Fodd101}) is equal to zero.
Further, the term $\partial_t^{(0)}\sigma_k^{(2n+1)}$ gives the
linear contribution:
\[
\partial_t^{(0)}\sigma_k^{(2n+1)}=\partial_t^{(0)}
b_n (-k^2 )^{n} (-k^2 )p_k =-\frac{5}{3}b_n (-k^2 )^{n+1} iku_k ,
\]
while the terms in the remaining sum contribute nonlinearly:
\[
\partial_t^{(2m+1)}\sigma_k^{(2(n-m))}=
a_{n-m}(-k^2 )^{n-m}ik\partial_t^{(2m+1)}u_k
=-a_{n-m}a_m (-k^2 )^{n+1}ik u_k .
\]
Substituting the last two expressions into (\ref{Fodd101}), we see
that it has just the same structure as the coefficient
$\sigma_k^{(2(n+1))}$ in (\ref{Fstructure101}). Thus, we obtain
the first recurrence equation:
\[
a_{n+1}=\frac{5}{3}b_n +\sum_{m=0}^{n}a_{n-m}a_m .
\]
Computing the coefficient $\sigma_k^{(2(n+1)+1)}$ by the same
pattern, we come to the second recurrence equation, and the
Chapman--Enskog procedure (\ref{procedure101}) and
(\ref{operators101}) can be reformulated in terms of the
coefficients $a_n $ and $b_n $ (\ref{structure101}):
\begin{eqnarray}
\label{recurrent101}
a_{n+1}&=&\frac{5}{3}b_n +\sum_{m=0}^{n}a_{n-m}a_m ,\\\nonumber
b_{n+1}&=&a_{n+1}+\sum_{m=0}^{n}a_{n-m}b_m .
\end{eqnarray}
The initial condition for this set of equations is dictated by the
Navier--Stokes and the Burnett terms:
\begin{equation}
\label{initialcondition101}
a_0 = -\frac{4}{3},\quad
b_0 =-\frac{4}{3}
\end{equation}

Our goal now is to compute the functions $A$ and $B$ (\ref{AB101})
on the basis of the recurrence equations (\ref{recurrent101}). At
this point, it is worthwhile to notice that the usual way of
dealing with the recurrence system (\ref{recurrent101}) would be
either to truncate it at a certain $n$, or to calculate all the
coefficients explicitly, and substitute the result into the power
series (\ref{AB101}). Both approaches are not successful here.
Indeed, retaining the coefficients $a_0$, $b_0$, and $a_1$ gives
the super-Burnett approximation (\ref{sburnett101}) which has the
Bobylev short-wave instability, and there is no guarantee that the
same failure will not occur in the higher-order truncation. On the
other hand, a term-by-term computation of the whole set of
coefficients $a_n$ and $b_n$ is a nontrivial task due to the
nonlinearity in (\ref{recurrent101}).

Fortunately, another route is possible.
Multiplying both the equations in (\ref{recurrent101}) with $(-k^2
)^{n+1}$, and performing a formal
summation in $n$ from zero to infinity, we
arrive at the following expressions:
\begin{eqnarray}
\label{intermediate101}
A-a_0& =&-k^2\left\{\frac{5}{3}B+\sum_{n=0}^{\infty}\sum_{m=0}^{n}
a_{n-m}(-k^2 )^{n-m}a_m (-k^2 )^{m}\right\}, \\\nonumber
B-b_0& =&A-a_0 -k^2\sum_{n=0}^{\infty}\sum_{m=0}^{n}
a_{n-m}(-k^2 )^{n-m}b_m (-k^2 )^{m} .
\end{eqnarray}
Now we notice that
\begin{eqnarray}
\lim_{N\rightarrow\infty}\sum_{n=0}^{N}\sum_{m=0}^{n}
a_{n-m}(-k^2 )^{n-m}a_m (-k^2 )^{m}&=&A^2, \\\nonumber
\lim_{N\rightarrow\infty}\sum_{n=0}^{N}\sum_{m=0}^{n}
a_{n-m}(-k^2 )^{n-m}b_m (-k^2 )^{m}&=&AB .
\end{eqnarray}
Taking into account the initial condition
(\ref{initialcondition101}), equation (\ref{intermediate101})
yields a pair of coupled quadratic equations for the functions $A$
and $B$:
\begin{eqnarray}
\label{system101}
A&=&-\frac{4}{3}-k^2 \left(\frac{5}{3}B+A^2 \right),\\\nonumber
B&=&A(1-k^2 B).
\end{eqnarray}

The result (\ref{system101}) concludes essentially the question of
the computation of functions $A$ and $B$ (\ref{AB101}). Still,
further simplifications are possible. In particular, it is
convenient to reduce the consideration to a single function.
Solving system (\ref{system101}) for $B$, and introducing a new
function,  $X(k^2 )=k^2 B(k^2 )$, we obtain an equivalent cubic
equation:
\begin{equation}
\label{factor101}
-\frac{5}{3}(X-1)^2 \left(X+\frac{4}{5}\right)=\frac{X}{k^2 }.
\end{equation}
Since $A$ and $B$ (\ref{AB101}) are real-valued, we are only
interested in the real-valued roots of (\ref{factor101}).

\begin{figure}[t]
\centering {
\includegraphics[width=100mm, height=80mm]{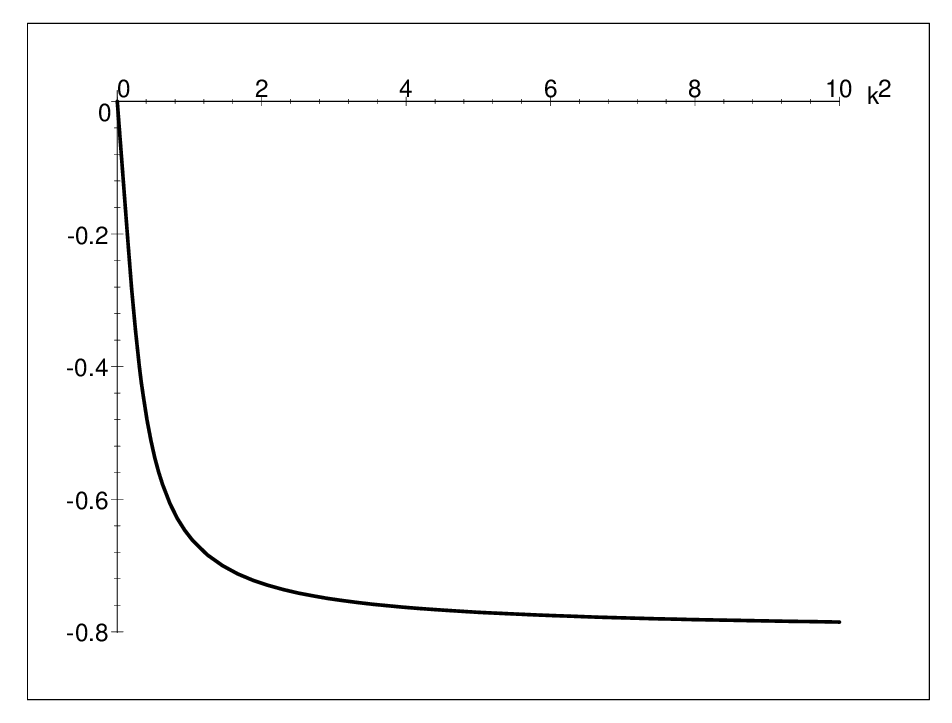}
\caption {\label{AnnPhysFig1}Real-valued root of (\ref{factor101}) as a function of $k^2$.}}
\end{figure}

An elementary analysis of this equation brings the following
result: {\it the real-valued root $X(k^2 )$ of (\ref{factor101})
is unique and {\it negative} for all finite values  $k^2 $}.
Moreover, the function $X(k^2 )$ is a monotonic function of $k^2 $
(Fig.~\ref{AnnPhysFig1}). The limiting values are:
\begin{equation}
\label{limits101}
\lim_{|k|\rightarrow0}X(k^2 )=0, \quad
\lim_{|k|\rightarrow\infty}X(k^2 )=-0.8.
\end{equation}

\begin{figure}[t]
\centering{
\includegraphics[width=100mm, height=80mm]{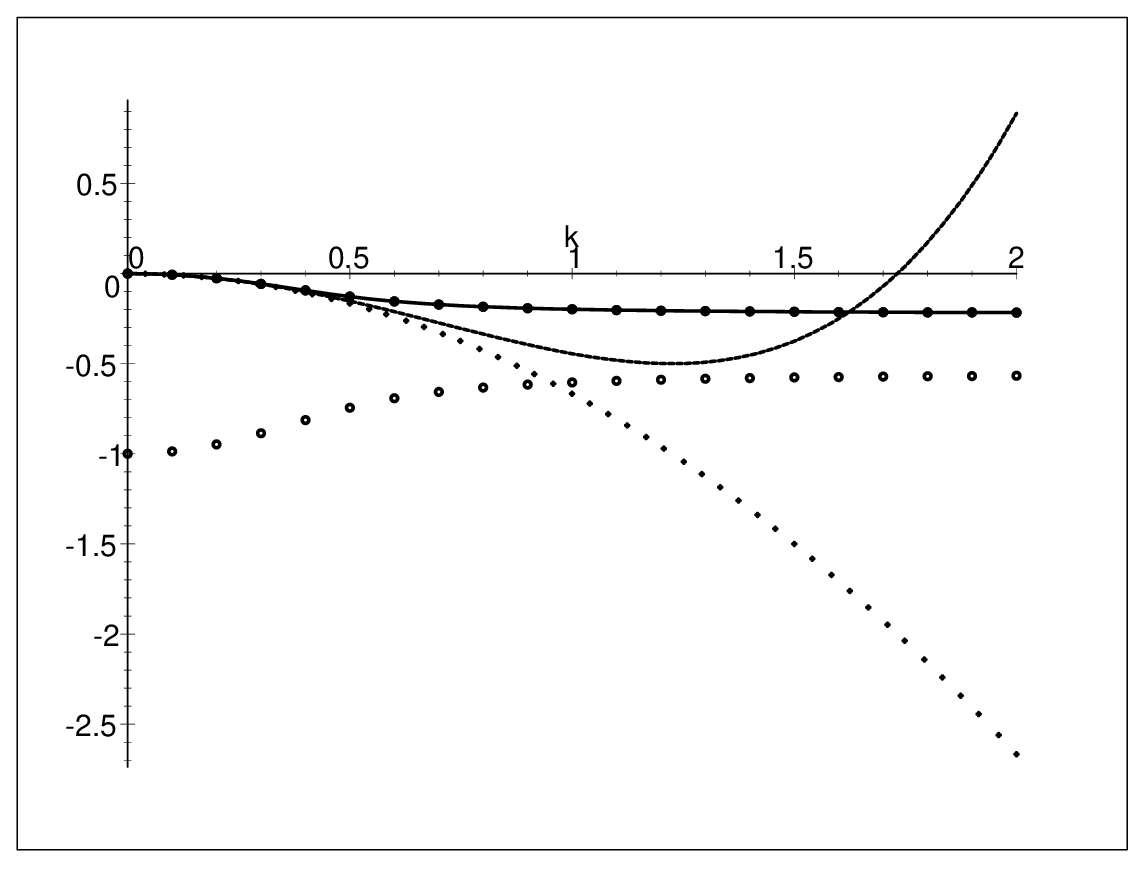}
\caption {\label{AnnPhysFig2}Attenuation rates for the $1D10M$
Grad system. Solid: Exact summation of the Chapman--Enskog
Expansion. Dots: The Navier--Stokes approximation. Dash: The
super--Burnett approximation. Circles: Hydrodynamic and
non-hydrodynamic modes of the $1D10M$ Grad system.}}
\end{figure}

Under the conditions just mentioned, the function under the root
in (\ref{dispersion101}) is negative for all values of the wave
vector $k$, including the limits, and we come to the following
dispersion law:
\begin{equation}
\label{frequency101}
\omega_{\pm}=\frac{X}{2(1-X)}\pm i\frac{|k|}{2}\sqrt{\frac{5X^2
-16X +20}{3}},
\end{equation}
where $X=X(k^2 )$ is the real-valued root of equation
(\ref{factor101}). Since $X(k^2 )$ is negative for all $|k|>0$,
the attenuation rate, ${\rm Re}( \omega_{\pm})$, is negative for
all $|k|>0$, and the exact acoustic spectrum of the
Chapman--Enskog procedure {\it is stable for arbitrary wave
lengths}. In the short-wave limit, from (\ref{frequency101}) we
obtain:
\begin{equation}
\label{limit101}
\lim_{|k|\rightarrow\infty}\omega_{\pm}=-\frac{2}{9}\pm
i|k|\sqrt{3}.
\end{equation}

The characteristic equation of the original Grad equations
(\ref{Grad101}) reads:
\begin{equation}
\label{dispersiongrad101}
3\omega^3 +3\omega^2 + 9k^2 \omega +5k^2 =0.
\end{equation}
The two complex-conjugate roots of this equation correspond to the
hydrodynamic modes, while for the non-hydrodynamic real mode,
$\omega_{nh}(k)$, $\omega_{nh}(0)=-1$, and $\omega_{nh}\rightarrow
-0.5$ as $|k|\rightarrow\infty$. Recall that the non-hydrodynamic
modes of the Grad equations are characterized by the common
property that for them $\omega(0)\ne 0$. These modes are
irrelevant to the Chapman--Enskog method. As the final comment
here, (\ref{limit101}) demonstrates that the exact attenuation
rate, ${\rm Re} ( \omega_{\pm})$, tends to a finite value,
$-\frac{2}{9}\approx-0.22$ as $|k|\rightarrow\infty$. This
asymptotic behavior is in a complete agreement with the data for
the hydrodynamic branch of the spectrum (\ref{dispersiongrad101})
of the original Grad equations (\ref{Grad101}). The attenuation
rates (real parts of the dispersion relations $\omega_{\pm}$ for
the Burnett (\ref{dispersionburnett101}), the super-Burnett
(\ref{dispersionsburnett101}), the exact Chapman--Enskog solution
(\ref{frequency101}), are compared to each other in
Fig.~\ref{AnnPhysFig2}. In this figure, we also represent the
attenuation rates of the hydrodynamic and non-hydrodynamic mode of
the  Grad equations (\ref{dispersiongrad101}). The results of this
section lead to the following conclusion:

(i) The proposed approach provides a way to deal with the problem
of {\it summation} of the Chapman--Enskog expansion. The exact
dispersion relation (\ref{frequency101}) of the Chapman--Enskog
procedure is demonstrated to be stable for all wave lengths, while
the Bobylev instability is present on the level of the
super-Burnett approximation. Moreover, it can be demonstrated that
the function $X$ (the real root of (\ref{factor101})) is a
real-valued analytic function of $k$. Thus, the treatment of the
formal expansions performed above is justified.

(ii) The exact result of the Chapman--Enskog procedure has a clear
non-polynomial character. Indeed, this follows directly from
(\ref{limits101}): the function $X(k^2 )$ cannot be a polynomial
because it maps the axis  $k $ into a segment $[0,-0.8]$. As a
conjecture here, the resulting exact hydrodynamics is {\it
essentially } nonlocal in space. For this reason, even if the
hydrodynamic equations of a certain level of the approximation
{\it is} stable, it cannot reproduce the non-polynomial behavior
for sufficiently short waves.

(iii) The result of this section demonstrates that, at least in
some cases, the sum of the Chapman--Enskog series amounts to a
quite regular function, and the ``smallness" of the Knudsen number
$\epsilon$ used to develop the Chapman--Enskog procedure
(\ref{procedure101}) {\it is no longer necessary}.

\subsection{The $3D10M$ Grad equations} \label{exact_10_3}

In this section we generalize our considerations of the
Chapman--Enskog method to the three-dimensional linearized
10-moment Grad equations \cite{Grad}. The Chapman--Enskog series
for the stress tensor, which is again due to a nonlinear
procedure, will be  summed up in closed form. The method used
follows essentially the one discussed above, though the
computations are slightly more extensive. The reason to consider
this example is that we would like to know what happens to the
diffusive hydrodynamic mode in the short-wave domain.

Throughout this section, we use the variables (\ref{AnnPhysvar}),
and $p$ and ${\bfu}$ are dimensionless deviations of pressure and
of mean flux from their equilibrium values, respectively. The
point of departure is the set of the three-dimensional linearized
Grad equations for the $p$, ${\bfu}$, and $\s $, where $\s $ is a
dimensionless stress tensor:
\begin{eqnarray}
\label{Grad103}
\partial_t p &=&-\frac{5}{3}\nabla \cdot {\bfu},\\\nonumber
\partial_t {\bfu} &=&-\nabla p -\nabla\cdot\s ,
\\\nonumber
\partial_t  \s &=&-\overline{\nabla{\bfu}}
-\frac{1}{\epsilon}\s.
\end{eqnarray}

Equation (\ref{Grad103}) provides a simple model of a coupling of
the hydrodynamic variables, ${\bfu}$ and $p$, to the
non-hydrodynamic variable ${\s}$. These equations are suitable for
an application of the Chapman--Enskog procedure. Therefore, our
goal here is not to investigate the properties of (\ref{Grad103})
as they are, but to reduce the description, and  to get a closed
set of equations with respect to the variables $p$ and ${\bfu}$
only. That is, we have to express  $\s $ in terms of spatial
derivatives of $p$ and of ${\bfu}$. The Chapman--Enskog method, as
applied to (\ref{Grad103}) results in the following:
\begin{equation}
\label{series103}
{\s}=\sum_{n=0}^{\infty}\epsilon^{n+1}{\s}^{(n)}.
\end{equation}
The coefficients $\s^{(n)}$ are due to the following recurrence
procedure:
\begin{equation}
\label{procedure103}
\s^{(n)}=-\sum_{m=0}^{n-1}\partial_t^{(m)}\s^{(n-1-m)},
\end{equation}
where the Chapman--Enskog operators $\partial_t^{(m)}$ act on the
functions $p$ and ${\bfu}$, and on their derivatives, as follows:
\begin{eqnarray}
\label{operators103}
\partial_t^{(m)}D{\bfu}&=&\left\{
\begin{array}{l@{\quad}l}
-D\nabla p, & m=0\\
-D\nabla \cdot \s^{(m-1)}, & m\ge 1\end{array}\right.,
\\\nonumber
\partial_t^{(m)}Dp&=&\left\{
\begin{array}{l@{\quad}l}
-\frac{5}{3}D\nabla\cdot{\bfu}, & m=0\\
0, & m\ge 1\end{array}\right..
\end{eqnarray}
Here $D$ is an arbitrary differential operator $D=\prod_{i=1}^3
\partial_i^{l_i} $, while $l_i$ is an arbitrary integer, and
$\partial_i^0 =1$. Finally, $\s^{(0)}=-\overline
{{\nabla}{\bfu}}$, which leads to the Navier--Stokes
approximation.

Our goal is to sum up the series (\ref{series103}) in closed form.

The terms  $\s^{(n)}$ in equations (\ref{series103}),
(\ref{procedure103}), and (\ref{operators103}), have the following
explicit structure for arbitrary order $n\ge 0$ (a generalization
of  (\ref{structure101}) to the three-dimensional case):
\begin{eqnarray}
\label{structure103}
\s^{(2n)}&=&a_n \Delta^n \overline {{\nabla}{\bfu}}+b_n
\Delta^{n-1} {G}{\nabla}\cdot{\bfu},\\\nonumber
\s^{(2n+1)}&=&c_n \Delta^n {G}p,
\end{eqnarray}
where $\Delta=\nabla\cdot\nabla$ is the Laplace operator, and the
operator ${G}$ has the form:
\begin{equation}
\label{operatorG}
{
G}=\nabla\nabla-\frac{1}{3}{I}\Delta=\frac{1}{2}\overline{\nabla\nabla}.
\end{equation}
The real-valued and yet unknown coefficients $a_n$, $b_n $, and
$c_n $ in  (\ref{structure103}) are due to the recurrence
procedure (\ref{procedure103}), and (\ref{operators103}). Knowing
the structure of the coefficients of the Chapman--Enskog series
(\ref{structure103}), we can reformulate the Chapman--Enskog
solution in terms of a self-consistent recurrence procedure for
the coefficients $a_n $, $b_n $, and $c_n $. Let us consider this
derivation in  more detail.

The point of departure is the Fourier representation of the
recurrence equations (\ref{procedure103}), (\ref{operators103}),
and (\ref{structure103}). Writing
\begin{eqnarray*} {\bfu}&=&{\bfu}_{k}\exp(i{\bfk}
\cdot{\bfx}),\\
p&=&p_{k}\exp(i{\bfk}
\cdot{\bfx}),\\
\s^{(n)}&=&\sk^{(n)}\exp(i{\bfk}
\cdot{\bfx}),
\end{eqnarray*}
and introducing the unit vector $\bfe\kkk$ directed along ${\bfk}$
(${\bfk}=k{\bfe}\kkk$), equations (\ref{procedure103}),
(\ref{operators103}), and (\ref{structure103}) can be rewritten
as:
\begin{equation}
\label{Fprocedure103}
\sk^{(n)}=-\sum_{m=0}^{n-1}\partial_t^{(m)}\sk^{(n-1-m)},
\end{equation}
\begin{eqnarray}
\label{Foperators103}
\partial_t^{(m)}D\kkk{\bfu}\kkk&=&\left\{
\begin{array}{l@{\quad}l}
-D\kkk i{\bfk} p\kkk, & m=0\\ -D\kkk i{\bfk} \cdot \sk^{(m-1)}, & m\ge 1\end{array}\right.,
\\\nonumber
\partial_t^{(m)}D\kkk p\kkk&=&\left\{
\begin{array}{l@{\quad}l}
-\frac{5}{3}D\kkk i{\bfk}\cdot{\bfu}\kkk, & m=0\\ 0, & m\ge 1\end{array}\right..
\end{eqnarray}
where $D\kkk $ is an arbitrary tensor $D\kkk=\prod_{s=1}^{3}(ik_s )^{l_s}$, and
\begin{eqnarray}
\label{Fstructure103} \sk^{(2n)}&=&(-k^2 )^{n}(a_n i \overline {{\bfk}{\bfu}}+b_n i {\bfg}\kkk
({\bfk}\cdot{\bfu})),\\\nonumber \sk^{(2n+1)}&=&c_n (-k^2 )^{n+1} {\bfg}\kkk p\kkk,
\end{eqnarray}
where
\begin{equation}
\label{FoperatorG} {\bfg}_{k}=({\bfe}\kkk{\bfe}\kkk-\frac{1}{3}{I}) =\frac{1}{2} \overline{
{\bfe}\kkk{\bfe}\kkk}.
\end{equation}

From the form of the Navier--Stokes
approximation, $\s_{k}^{(0)}$, it follows that $a_0 =-1$ and
$b_0 =0$, while a direct computation of the Burnett approximation leads
to:
\begin{equation}
\label{Fburnett103}
\s_{k}^{(1)}= \frac{1}{2}k^2 {\bfg}_{k}p_{k}.
\end{equation}
Thus, we have $c_0 = -\frac{1}{2}$which proves the ansatz
(\ref{structure103}) for $n=0$ in both the even and the odd
orders.

The rest of the proof relies on induction. Let the structure
(\ref{Fstructure103}) be proven up to the   order $n$. The
computation of the next, $n+1$ order coefficient
$\s_{k}^{(2(n+1))}$, involves only terms of  lower order. From
(\ref{Fprocedure103}) we obtain:
\begin{equation}
\label{Fodd103}
\s_{k}^{(2(n+1))}=-\partial_t^{(0)}\s_{k}^{(2n+1)}-
\sum_{{m=1}}^{{2n+1}}\partial_t^{(m)}\s_{k}^{(2n+1-m)}.
\end{equation}
The first term in the right hand side depends linearly on the
coefficients $c_{n}$:
\begin{eqnarray}
\label{Flinearodd103} -\partial_t^{(0)}\s_{k}^{(2n+1)}&=& -c_n (-k^2 )^{n+1}{\bfg}\kkk
\partial_t^{(0)}p\kkk \\\nonumber &=& \frac{5}{3}c_n (-k^2 )^{n+1}i {\bfg}_{k}{\bfk}\cdot{\bfu}_{k}.
\end{eqnarray}
The remaining terms on the right hand side of (\ref{Fodd103})
contribute nonlinearly. Splitting the even and the odd orders of
the Chapman--Enskog operators $\partial_t^{(m)}$, we rewrite the
sum in  (\ref{Fodd103}):
\begin{equation}
\label{Foddelement0103}
-\sum_{{m=1}}^{{2n+1}}\partial_t^{(m)}\s_{k}^{(2n+1-m)}=
-\sum_{{l=1}}^{{n}}\partial_t^{(2l)}\s_{k}^{(2(n-l)+1)}
-\sum_{{l=0}}^{{n}}\partial_t^{(2l+1)}\s_{k}^{(2(n-l))}.
\end{equation}
Due to (\ref{Fstructure103}) and (\ref{Foperators103}), each term
in the first sum is equal to zero, and we are left only with the
second sum:
\begin{equation}
\label{Foddelement1103}
\partial_t^{(2l+1)}\s_{k}^{(2(n-l))}=(-k^2 )^{n-l}
(a_{n-l}i\overline{{\bfk}\partial_t^{(2l+1)}{\bfu}\kkk }+ b_{n-l}i{\bfg}\kkk
{\bfk}\cdot\partial_t^{(2l+1)}{\bfu}\kkk ),
\end{equation}
while
\begin{equation}
\label{Foddelement2103}
\partial_t^{(2l+1)}{\bfu}\kkk =-(-k^2 )^{l+1}(a_l {\bfu}\kkk +
\frac{1}{3}(a_l + 2 b_l ){\bfe}\kkk ({\bfe}\kkk \cdot {\bfu}\kkk )).
\end{equation}
In the last expression, use of the following identities was made:
\begin{eqnarray}
\label{Fid} {\bfk}\cdot \overline{{\bfk}{\bfu}\kkk }&=&k^2 ({\bfu}\kkk +\frac{1}{3}{\bfe}\kkk
({\bfe}\kkk \cdot {\bfu}\kkk )), \\\nonumber {\bfk}\cdot{\bfg}\kkk &=&\frac{2}{3}{\bfk}.
\end{eqnarray}
Substituting (\ref{Foddelement2103}) into the right hand side of
(\ref{Foddelement1103}), and thereafter substituting the result
into the right hand side of (\ref{Foddelement0103}), we obtain the
following in the right hand side of (\ref{Fodd103}):
\begin{eqnarray}
&&\s_{k}^{(2(n+1))}= (-k^2)^{n+1}\left(\sum_{m=0}^{n}a_{n-m}a_m \right) i\overline{{\bfk} {\bfu}_{k}}
+(-k^2 )^{n+1}\left(\frac{5}{3}c_n \right. \\\nonumber && \left. \; +
\sum_{m=0}^{n}\left\{\frac{1}{3}(2a_{n-m}+b_{n-m})(a_m + 2b_m )+a_{n-m}b_m
\right\}\right)i{\bfg}_{k}({\bfk}\cdot{\bfu}_{k}).
\end{eqnarray}
The functional structure of the right hand side of this expression
is  the same as that of the first  equation in the set
(\ref{Fstructure103}), and thus we obtain the first recurrence
equation:
\begin{eqnarray}
\label{firstabc103}&& a_{n+1}\overline{{\bfk}{\bfu}_{k}} + b_{n+1}{\bfg}_{k} ({\bfk}\cdot{\bfu}_{k})
=
\left( \sum_{m=0}^{n}a_{n-m}a_m \right) \overline{{\bfk} {\bfu}_{k}}\\ \nonumber&& \; +
\left(\frac{5}{3}c_n + \sum_{m=0}^{n}\left\{\frac{1}{3}(2a_{n-m}+b_{n-m})(a_m + 2b_m )+a_{n-m}b_m
\right\}\right){\bfg}_{k}({\bfk}\cdot{\bfu}_{k}).
\end{eqnarray}
Considering in the same way the coefficient $\s_{k}^{(2(n+1)+1)}$,
we come to the second recurrence equation,

\begin{equation}
\label{abc103-2}
c_{n+1}=2a_{n+1} +
b_{n+1}+\frac{2}{3}\sum_{m=0}^{n}(2a_{n-m}+b_{n-m})c_m .
\end{equation}
Thus, the complete set of the recurrence equations is given by  (\ref{firstabc103}) and
(\ref{abc103-2}). Equation (\ref{firstabc103}) is equivalent to a pair of scalar equations. Indeed,
introducing new variables,
\begin{eqnarray}
\label{var103}
r_{n}&=&\frac{2}{3}c_n ,\\\nonumber
q_n &=& \frac{2}{3}(2a_n + b_n ),
\end{eqnarray}
and using the identity, \[\overline{{\bfk}{\bfu}_{k}}=(
\overline{{\bfk}{\bfu}_{k}}-2{\bfg}_{k}({\bfk}\cdot{\bfu}
_{k}))+2{\bfg}_{k}({\bfk}\cdot{\bfu} _{k}),\] and also noticing
that \[ {\bfg}_{k}:(\overline{{\bfk}{\bfu}
_{k}}-2{\bfg}_{k}({\bfk}\cdot{\bfu} _{k}))=0,\] where $:$ denotes
the double contraction of tensors, we arrive in
(\ref{firstabc103}) and (\ref{abc103-2}) at the following three
scalar recurrence relations in terms the coefficients $r_n $, $q_n
$, and $a_n $:
\begin{eqnarray}
\label{arq103}
r_{n+1}&=&q_{n+1} +\sum_{m=0}^{n}q_{n-m}r_m \\\nonumber
q_{n+1}&=&\frac{5}{3}r_n +\sum_{m=0}^{n}q_{n-m}q_m \\\nonumber
a_{n+1}&=&\sum_{m=0}^{n}a_{n-m}a_m
\end{eqnarray}
The initial condition for this system is provided by the explicit
form of the Navier--Stokes and the Burnett approximations, and
reads:
\begin{equation}
\label{incondition103}
r_0 = -4/3, \quad q_0 =
-4/3,\quad a_0 = -1.
\end{equation}

The recurrence relations (\ref{arq103}) are completely equivalent
to the original Chapman--Enskog procedure (\ref{procedure103}) and
(\ref{operators103}). In the one-dimensional case, the recurrence
system (\ref{arq103}) reduces to  the first two equations for
$r_{n}$ and $q_n $. In this case, the system of recurrence
equations is identical (up to the notations) to the recurrence
system (\ref{recurrent101}), considered in the preceding section.
For what follows, it is important to notice that the recurrence
equation for the coefficients $a_n$ is decoupled from the
equations for the coefficients $r_n$ and $q_n$.

Now we shall express the Chapman--Enskog series of the stress
tensor (\ref{series103}) in terms of  $r_n $, $q_n $, and $a_n $.
Using again the Fourier transform, and substituting
(\ref{structure103}) into the right hand side of
(\ref{series103}), we derive:
\begin{equation}
\label{sigma103}
{\s}_{k}=A(k^2 )(\overline{{\bfk}{\bfu}
_{k}}-2{\bfg}_{k}({\bfk}\cdot{\bfu}
_{k}))+\frac{3}{2}Q(k^2 ){\bfg}_{k}({\bfk}\cdot{\bfu}
_{k})-\frac{3}{2}k^2 R(k^2 ){\bfg}_{k}p_{k},
\end{equation}
From here on, we use a new spatial scale which amounts to ${\bfk}'
=\epsilon {\bfk}$, and drop the prime. The functions $A(k^2 )$,
$Q(k^2 )$, and $R(k^2 )$ in (\ref{sigma103}) are defined by the
power series with the coefficients due to (\ref{arq103}):
\begin{eqnarray}
\label{ARQ103}
A(k^2 )&=&\sum_{n=0}^{\infty}a_n (-k^2 )^n , \\\nonumber
Q(k^2 )&=&\sum_{n=0}^{\infty}q_n (-k^2 )^n , \\\nonumber
R(k^2 )&=&\sum_{n=0}^{\infty}r_n (-k^2 )^n .
\end{eqnarray}

Thus, the question of summation of the Chapman--Enskog series
(\ref{series103}) amounts to finding the three functions, $A=A(k^2
)$,   $Q=Q(k^2 )$, and $R=R(k^2 )$ (\ref{ARQ103}) in the three-
and two-dimensional cases, or to the two functions, $Q(k^2 )$, and
$R(k^2 )$ in the one-dimensional case.

Now we shall focus on computing the functions (\ref{ARQ103}) from
the recurrence equations (\ref{arq103}). At this point, it is
worthwhile to notice again that a truncation  at a certain $n$ is
not successful. Indeed, already in the one-dimensional case,
retaining the coefficients $q_0$, $r_0$, and $q_1$ leads to the
super-Burnett approximation (\ref{sburnett101}) which has the
short-wave instability for $k^2
>3$, as it was demonstrated in the preceding section, and there is
no guarantee that the same will not occur in a higher-order
truncation.

Fortunately, the approach introduced in the preceding section
works again. Multiplying each of  the equations in (\ref{ARQ103})
with $(-k^2 )^{n+1}$, and performing a summation in $n$ from zero
to infinity, we derive:
\begin{eqnarray}
\label{intermediate103}
Q-q_0& =&-k^2\left\{\frac{5}{3}R+\sum_{n=0}^{\infty}\sum_{m=0}^{n}
q_{n-m}(-k^2 )^{n-m}q_m (-k^2 )^{m}\right\}, \\\nonumber
R-r_0& =&Q-q_0 -k^2\sum_{n=0}^{\infty}\sum_{m=0}^{n}
q_{n-m}(-k^2 )^{n-m}r_m (-k^2 )^{m},
\\\nonumber
A-a_0&=&-k^2 \sum_{n=0}^{\infty}\sum_{m=0}^{n}
a_{n-m}(-k^2 )^{n-m}a_m (-k^2 )^{m}.
\end{eqnarray}
Now we notice that
\begin{eqnarray}
\label{trick103}
\lim_{N\rightarrow\infty}\sum_{n=0}^{N}\sum_{m=0}^{n}
a_{n-m}(-k^2 )^{n-m}a_m (-k^2 )^{m}&=&A^2 , \\\nonumber
\lim_{N\rightarrow\infty}\sum_{n=0}^{N}\sum_{m=0}^{n}
q_{n-m}(-k^2 )^{n-m}r_m (-k^2 )^{m}&=&QR , \\\nonumber
\lim_{N\rightarrow\infty}\sum_{n=0}^{N}\sum_{m=0}^{n}
q_{n-m}(-k^2 )^{n-m}q_m (-k^2 )^{m}&=&Q^2 .
\end{eqnarray}
Taking into account  the initial conditions
(\ref{incondition103}), and also using (\ref{trick103}), we derive
from (\ref{intermediate103}) the following three quadratic
equations for the functions $A$, $R$, and $Q$:
\begin{eqnarray}
\label{system103}
Q&=&-\frac{4}{3}-k^2 \left(\frac{5}{3}R+Q^2 \right),\\\nonumber
R&=&Q(1-k^2 R), \\\nonumber
A&=&-(1+k^2 A^2 ).
\end{eqnarray}
The result (\ref{system103}) concludes essentially the question of
computation of functions (\ref{ARQ103}) in closed form. Still,
further simplifications are possible. In particular, it is
convenient to use a single unknown function , $X(k^2 )=k^2 R(k^2
)$, in the first two equations in the system (\ref{system103}). We
again obtain an equivalent cubic equation:
\begin{equation}
\label{factor103} -\frac{5}{3}(X-1)^2 (X+\frac{4}{5})=\frac{X}{k^2
},
\end{equation}
which coincides with  (\ref{factor103}) of the previous section.
We shall also rewrite the third equation of (\ref{system103})
using a function $Y(k^2 )=k^2 A(k^2 )$:
\begin{equation}
\label{quadratic103}
Y(1+Y)=-k^2 .
\end{equation}

The functions in (\ref{ARQ103}) can now be straightforwardly
expressed in terms of the  relevant solutions to (\ref{factor103})
and (\ref{quadratic103}). Since all functions in (\ref{ARQ103})
are real-valued functions, we are interested only in the
real-valued roots of the algebraic equations (\ref{factor103}) and
(\ref{quadratic103}).

The relevant analysis of the cubic equation (\ref{factor103}) was
already performed above: the real-valued root $X(k^2 )$ is unique
and negative for all finite values of $k^2 $. Limiting values of
the function  $X(k^2 )$ at $k\rightarrow 0$ and at $k\rightarrow
\infty$ are given by (\ref{limits101}):
\[
\lim_{k\rightarrow0}X(k^2 )=0, \quad
\lim_{k\rightarrow\infty}X(k^2 )=-\frac{4}{5}.
\]

The quadratic equation (\ref{quadratic103}) has no real-valued
solutions for $k^2 >\frac{1}{4}$, and it has two real-valued
solution for each $k^2 $, where $k^2 <\frac{1}{4}$. We denote
$k_{\rm c} =\frac{1}{2}$ the corresponding critical value of the
wave vector. For $k =0$, one of these roots is equal to zero,
while the other is equal to one. The asymptotics $Y\rightarrow 0$,
as $k \rightarrow 0$, answers the question which of these two
roots of (\ref{quadratic103}) is relevant to the Chapman--Enskog
solution, and we derive:
\begin{equation}
\label{root103}
Y=\left\{ \begin{array}{ll}-\frac{1}{2}\left(1-\sqrt{1-
4k^2 }\right) & k <k_{\rm c} \\
{\rm none} & k > k_{\rm c} \end{array}\right.
\end{equation}
The function $Y$ (\ref{root103}) is negative for $k \le k_{\rm c} $.

From now on,  $X$ and $Y$ will denote the relevant roots of
(\ref{factor103}) and (\ref{quadratic103}) just discussed. The
Fourier image of the expression $\nabla\cdot{\s}$ follows from
(\ref{sigma103}):
\begin{equation}
\label{gradientsigma103}
i{\bfk}\cdot{\s}_{k}=Y(({\bfe}_{k}\cdot{\bfu}_{k})
{\bfe}_{k}-{\bfu}
_{k})-\frac{X}{1-X}({\bfe}_{k}\cdot{\bfu}_{k})
{\bfe}_{k}
-iX{\bfk}p_{k}.
\end{equation}
The latter expression contributes to the right-hand side of the
second of equations in the Grad system (\ref{Grad103}) (more
specifically, it contributes to the corresponding Fourier
transform of this equation). Knowing (\ref{gradientsigma103}), we
can calculate the dispersion $\omega({\bfk})$ of the plane waves
$\sim \exp\{\omega t + i{\bfk}\cdot{\bfx}\}$ which now follows
from the exact solution of the Chapman--Enskog procedure. The
calculation of the dispersion relation amounts to an evaluation of
the determinant of a $(d+1)\times(d+1)$ matrix, and is quite
standard (see, e.g. \cite{Resibois}). We therefore provide only
the final result. The exact dispersion relation of the
hydrodynamic modes reads:
\begin{equation}
\label{dispersion103}
\left(\omega - Y\right)^{d-1}\left(\omega^2 -\frac{X}{1-X}\omega +
\frac{5}{3}k^2(1- X)\right)=0.
\end{equation}
Here, $d$ is the spatial dimension.

From the dispersion relation (\ref{dispersion103}), we easily derive the
following classification of the hydrodynamic modes:

(i) For $d=1$, the spectrum of the hydrodynamic modes is purely
acoustic with the dispersion $\omega_{\rm a}$ which is given by
(\ref{frequency101}):
\begin{equation}
\label{frequency103}
\omega_{\rm a}=\frac{X}{2(1-X)}\pm i\frac{k}{2}\sqrt{\frac{5X^2
-16X +20}{3}},
\end{equation}
where $X=X(k^2 )$ is the real-valued root of (\ref{factor103}).
Since $X$ is a negative function for all $k>0$, the attenuation
rate of the acoustic modes, ${\rm Re} ( \omega_{\rm a})$, is
negative for all $k>0$, and the exact acoustic spectrum of the
Chapman--Enskog procedure
 is
free of the Bobylev instability for arbitrary wave lengths.

(ii) For  $d>1$, the dispersion of the acoustic modes  is given by
 (\ref{frequency103}). As follows from the
Chapman--Enskog procedure, the diffusion-like (real-valued) mode
has the dispersion $\omega_{\rm d}$:
\begin{equation}
\label{diffusion103}
\omega_{\rm d}=\left\{
\begin{array}{ll}-\frac{1}{2}\left(1-\sqrt{1-
4k^2 }\right) & k <k_{\rm c}\\
{\rm none} & k > k_{\rm c} \end{array}\right.
\end{equation}
The diffusion mode is $(d-1)$ times degenerated, the corresponding
attenuation rate is negative for $k<k_{\rm c}$, and this mode {\it
cannot be extended beyond the critical value $k_{\rm c}
=\frac{1}{2}$  within the Chapman--Enskog method.}

The reason why this rather remarkable peculiarity of the
Chapman--Enskog procedure occurs can be  found upon closer
investigation of the spectrum of the underlying Grad moment system
(\ref{Grad103}).

Indeed, in the original system (\ref{Grad103}), besides the
hydrodynamic modes, there exist several non-hydrodynamic modes
which are irrelevant to the Chapman--Enskog solution. All these
non-hydrodynamic modes are characterized by the property that the
corresponding dispersion relations $\omega({\bfk})$ do not go to
zero, as $ k\rightarrow 0$. At the point $k_{\rm c} =\frac{1}{2}$,
the diffusion branch (\ref{diffusion103}) intersects with one of
the  non-hydrodynamic branches of (\ref{Grad103}). For larger
values of the wave vector $k$, these two branches produce a pair
of complex conjugate solutions with the real part equal to
$-\frac{1}{2}$. Thus, though the spectrum of the original
equations (\ref{Grad103})indeed continues past $k_{\rm c} $, {\it
the Chapman--Enskog method does not recognize this extension as
part of the hydrodynamic branch.} It is also interesting to notice
that if we would accept all the roots of (\ref{quadratic103}),
including the complex-values for $k > k_{\rm c} $, and not only
the real-valued root as suggested by the asymptotics of the
Chapman--Enskog solution (see the explanations preceding
(\ref{root103})), then we would come in (\ref{dispersion103}) to
the structure of the dispersion relation just mentioned.

\begin{figure}[t]
\centering{
\includegraphics[width=100mm, height=80mm]{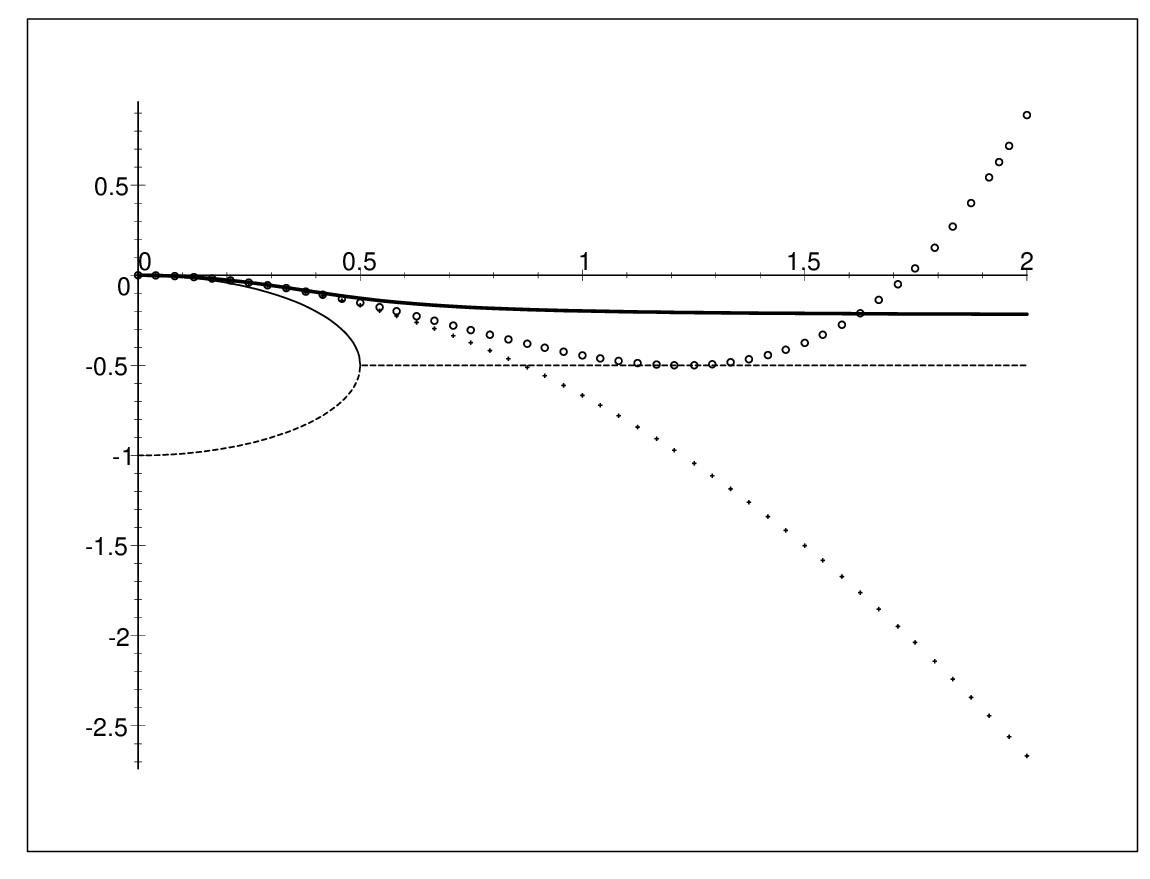}
\caption {\label{AnnPhysFig3} Attenuation rates for the $3D10M$
Grad system as functions of $|k|$. Bold: The acoustic branch,
exact summation. Dots: The acoustic branch, Navier--Stokes
approximation. Circles: The acoustic branch, super-Burnett
approximation. Solid: The diffusion branch, exact summation. Dash:
The critical mode of the $3D10M$ Grad system.}}
\end{figure}

The attenuation rates (the functions ${\rm Re}(\omega_{\rm a})$
and ${\rm Re}(\omega_{\rm d})$) are plotted in
Fig.~\ref{AnnPhysFig3}, together with the relevant dependencies
for the approximations of the Chapman--Enskog method. The
non-hydrodynamic branch of (\ref{Grad103}) which causes the
breakdown of the Chapman--Enskog solution is also represented in
Fig.~\ref{AnnPhysFig3}. It is rather remarkable that while the
exact hydrodynamic description becomes inapplicable for the
diffusion branch at $k\ge k_{\rm c} $, the usual Navier--Stokes
description still provides a good approximation to the acoustic
mode around this point.

The analysis of this section leads to the following additional
remarks to the conclusions made at the end of  Sect.
\ref{exact_10_1}:

(i) The developed approach provides an understanding of the
features of  Chapman--Enskog solutions and the problem of
extending the hydrodynamic modes into a highly non-equilibrium
domain on the exact basis and in the full spatial dimension. The
exact acoustic mode in  the framework of the Chapman--Enskog
procedure is demonstrated to be stable for all wave lengths, while
the diffusion-like mode can be regarded for the hydrodynamic mode
only in a bounded domain $k<k_{\rm c} $. It  is remarkable that
the result of the Chapman--Enskog procedure has a  clear
non-polynomial character. As a conjecture here, the resulting
hydrodynamics is {\it essentially } nonlocal in space. It is also
clear that {\it any} polynomial approximation to the
Chapman--Enskog series will fail to reproduce the peculiarity of
the diffusion mode demonstrated in the framework of the exact
solution.

(ii) Concerning the extension of hydrodynamics into a highly
non-equi\-li\-brium domain on  the basis of the Boltzmann
equations, the question remains open in the sense of an exact
summation as above. In this respect, results for simplified models
can serve either for testing  approximate procedures or at least
as guide. In particular, the mechanism of the singularity of the
diffusion-like mode through a coupling to the non-hydrodynamic
mode might be a rather general mechanism of limiting of the
hydrodynamic description, and not just a feature of the Grad
systems.

(iii) The result of this section demonstrates that the sum of the
Chapman--Enskog series amounts to either a quite regular function
(as is the function $X$), or to a function with a singularity at
finite $k_{\rm c}$. In both cases, however, the ``smallness" of
the Knudsen number $\epsilon$ used to develop the Chapman--Enskog
procedure plays no role in the result of the Chapman--Enskog
procedure.

\section{The dynamic invariance principle} \label{DI}
\subsection{Partial summation of the Chapman--Enskog expansion} \label{PS}

The examples considered above demonstrate that it makes sense to
speak about the  sum of the Chapman--Enskog expansion, at least
when the Chap\-man--Enskog method is applied to the (linearized)
Grad equations. However, even in this case, the possibility to
perform the summation exactly seems to be the lucky exception
rather than the rule. Indeed, computations become more bulky with
the increase of the number of the moments included in the Grad
equations. Therefore, we arrive at the question: how can we
approximate the recurrence equations of the Chapman--Enskog method
to account for all the orders in the Knudsen number? Any such
method amounts to some ``partial" summation of the Chapman--Enskog
expansion, and this type of working with formal series is widely
spread in various fields of physics.

In this section we shall discuss a method of approximating the
Chapman--Enskog expansion as a whole. As we now have the exact
expressions for the Chapman--Enskog solution for the linearized 10
moment Grad equations, it is natural to start with this example
for comparison purposes.

Let us come back to the originating one-dimensional Grad equations
(\ref{Grad101}), and to the corresponding formulas of the
Chapman--Enskog method (\ref{procedure101}) and
(\ref{operators101}). Instead of using the exact equations
(\ref{procedure101}) in each order $n$, we introduce the following
approximate equations:

Let $N \ge 1$ be some fixed integer. Then, instead of equations
(\ref{procedure101}), we write:
\begin{eqnarray}
\label{regprocedure101a}
\sigma^{(n)}&=&
-\sum_{m=0}^{n-1}\partial_t^{(m)}\sigma^{(n-1-m)}, \quad n\le N,
\\
\label{regprocedure101b}
\sigma^{(n)}&=&
-\sum_{m=0}^{N -1}\partial_t^{(m)}\sigma^{(n-1-m)}, \quad n>N.
\end{eqnarray}
This approximation amounts to the following: up to order $N$, the
Chapman--Enskog procedure (\ref{procedure101}) is taken exactly
(equation (\ref{regprocedure101a})), while in the computation of
higher orders (equation (\ref{regprocedure101b})) we restrict the
set of the Chapman--Enskog operators (\ref{operators101}) only up
to order $N$. Thus, the Chapman--Enskog coefficients
$\sigma^{(n)}$ of order higher than $N$ are taken into account
only ``partially". As $N$ tends to infinity, the recurrence
procedure (\ref{regprocedure101a}) and (\ref{regprocedure101b})
tends formally to the exact Chapman--Enskog procedure
(\ref{procedure101}). We shall further refer to
(\ref{regprocedure101a}) and (\ref{regprocedure101b}) as the {\it
regularization} of the $N-th$ order. In particular, taking $N=1$,
we come to the regularization of the Burnett approximation, taking
$N=2$ we come to the regularization of the super-Burnett
approximation, etc.

It can be demonstrated that the approximate procedure just
described does not alter the structure of the functions
$\sigma^{(2n)}$ and $\sigma^{(2n+1)}$ (\ref{structure101}), while
the recurrence equations for the coefficients $a_n $ and $b_n $
(\ref{structure101}) will differ from the exact result of the full
Chapman--Enskog procedure (\ref{recurrent101}). The advantage of
the regularization procedure (\ref{regprocedure101a}) and
(\ref{regprocedure101b}) over the exact Chapman--Enskog recurrence
procedure (\ref{procedure101}) is that the resulting equations for
the  coefficients $a_n $ and $b_n $ are always linear, as they
result from  (\ref{regprocedure101a}) and
(\ref{regprocedure101b}). This feature enables one to sum up the
corresponding series exactly, even if the originating nonlinear
procedure leads to a too difficult analysis. The number $N$ can be
called the ``depth" of the approximation: the large  $N$ is, the
more low-order terms of the Chapman--Enskog expansion are taken
into account exactly due to (\ref{regprocedure101a}).

For the first example, let us take $N=1$ in
(\ref{regprocedure101a}) and  (\ref{regprocedure101b}). The
regularization of the Burnett approximation then reads:
\begin{equation}
\label{regburnett101}
\sigma^{(n)}=-\partial_t^{(0)}\sigma^{(n-1)},
\end{equation}
where $n\ge1$, and $\sigma^{(0)}=-(4/3)\partial_x u$. Turning to the
Fourier variables, we derive:
\begin{eqnarray}
\label{regFstructure101}
\sigma_k^{(2n)}&=&a_n (-k^2 )^n ik u_k ,\\\nonumber
\sigma_k^{(2n+1)}&=&b_n (-k^2 )^{n+1}  p_k ,
\end{eqnarray}
where the coefficients $a_n $ and $b_n $ are due to the following
recurrence procedure:
\begin{equation}
\label{reg1recurrent101}
a_{n+1}=\frac{5}{3}b_n ,\quad b_n = a_n , \quad a_0 =-\frac{4}{3},
\end{equation}
whereupon
\begin{equation}\label{reg101}
a_n = b_n =\left(\frac{5}{3}\right)^n a_0 .
\end{equation}
Thus, denoting as $\sigma_{1k}^R $ the Fourier transform of the
regularized Burnett approximation, we obtain:
\begin{equation}
\label{1R101}
\sigma_{1k}^R = -\frac{4}{3+5k^2 }\left( iku_k -k^2 p_k \right).
\end{equation}

It should be noted that the recurrence equations
(\ref{reg1recurrent101}) can also be obtained from the exact
recurrence equations (\ref{recurrent101}) by neglecting the
nonlinear terms. Thus, the approximation adopted within the
regularization procedure (\ref{regburnett101}) amounts to the
following rational approximation of the functions $A$ and $B$
(\ref{AB101}):
\begin{equation}
\label{reg1AB101}
A_1^R = B_1^R =-\frac{4}{3+5k^2 }.
\end{equation}
Substituting the latter expressions instead of the functions $A$
and $B$ in the dispersion formula  (\ref{dispersion101}), we come
to the dispersion relation of the hydrodynamic modes within the
regularized Burnett approximation:
\begin{equation}
\label{reg1dispersion101}
\omega_{\pm}=-\frac{2k^2 }{3+5k^2 } \pm i|k|\sqrt{\frac{75k^2 k^2
+66k^2 + 15}{25 k^2 k^2 + 30 k^2 +9}}.
\end{equation}
The dispersion relation (\ref{reg1dispersion101}) is stable for all
wave vectors, and in the short-wave limit we have:
\begin{equation}
\label{reg1limit101}
\lim_{|k|\rightarrow \infty}\omega_{\pm}=-0.4 \pm i|k|\sqrt{3}.
\end{equation}

Thus, the regularized Burnett approximation leads  qualitatively
to the same behavior of the dispersion relation, as the exact
result (\ref{limit101}), with the limiting value of the
attenuation rate equal to $-0.4$ instead of the exact value
$-2/9$.

Consider now the regularization of the super-Burnett
approximation. This amounts to setting $N=2$ in the recurrence
equations (\ref{regprocedure101a}) and (\ref{regprocedure101b}).
Then, instead of (\ref{regburnett101}), we have:
\begin{eqnarray}
\label{regsburnett101}
\sigma^{(1)}&=&-\partial_t^{(0)}\sigma^{(0)}, \\\nonumber
\sigma^{(2+n)}&=&-\partial_t^{(0)}\sigma^{(n+1)}-
\partial_t^{(1)}\sigma^{(n)},
\end{eqnarray}
where $n\ge 0$. The corresponding recurrence equations for the
coefficients $a_n $ and $b_n $ now become:
\begin{equation}
\label{reg2recurrent101}
a_{n+1}=\frac{1}{3}b_n ,\quad a_n =b_n , \quad a_0 =-\frac{4}{3}.
\end{equation}
Thus, instead of (\ref{reg1AB101}), we obtain:
\begin{equation}
\label{reg2AB101}
A_2^R = B_2^R =-\frac{4}{3+k^2 }.
\end{equation}
The corresponding dispersion relation of the regularized super-Burnett
approximation reads:
\begin{equation}
\label{reg2dispersion101}
\omega_{\pm}=-\frac{2k^2 }{3+k^2 }\pm i|k|\sqrt{\frac{25k^2 k^2 + 78
k^2 +45}{3k^2 k^2 +18k^2 +27}},
\end{equation}
while in the short-wave limit the asymptotic behavior becomes:
\begin{equation}\label{reg2limit101}
\lim_{|k|\rightarrow \infty}\omega_{\pm}=-2 \pm i|k|\sqrt{\frac{25}{3}}.
\end{equation}

\begin{figure}[t]
\centering{
\includegraphics[width=100mm, height=80mm]{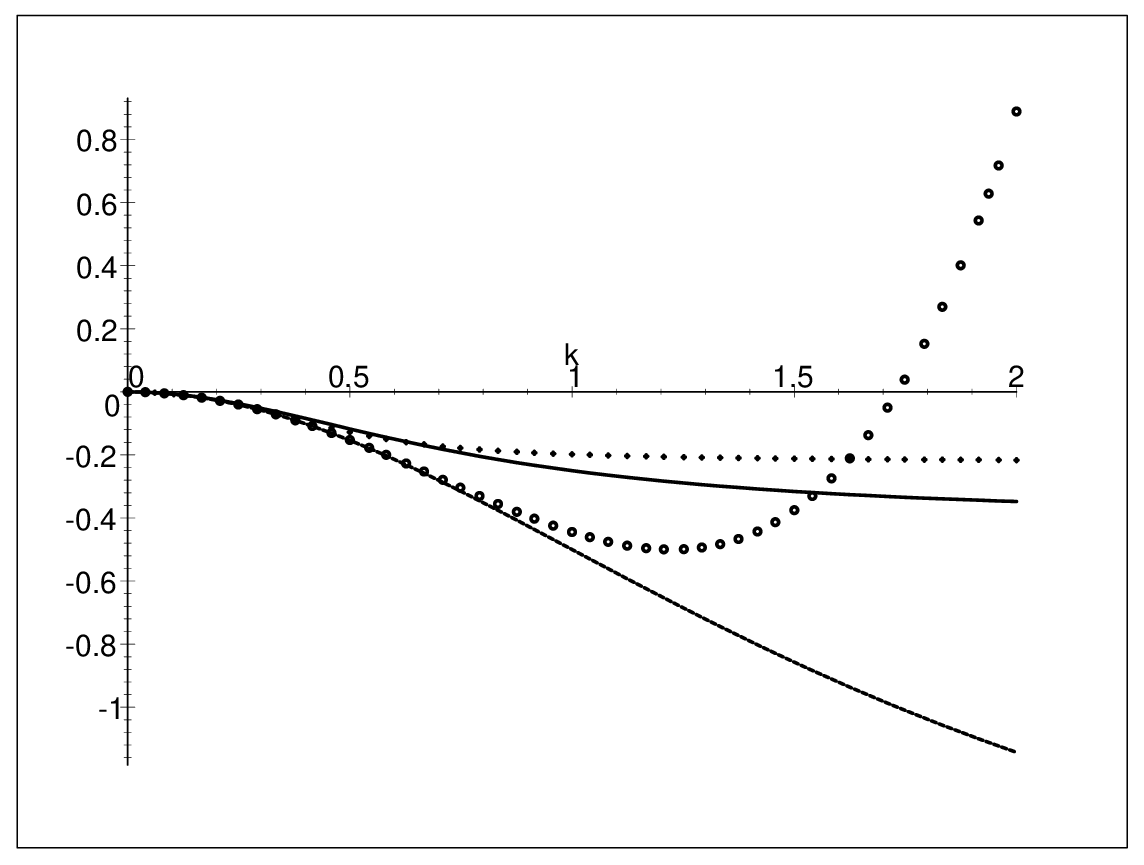}
\caption{\label{AnnPhysFig4} Attenuation rates for the partial summing. Solid: The regularized
Burnett approximation. Dash: The regularized super-Burnett approximation. Circles: The super-Burnett
approximation. Dots: The exact summation.}}
\end{figure}

The Bobylev instability is removed again within the regularization
of the super-Burnett approximation, and the lower-order terms of
the Chapman--Enskog expansion are taken into account more
precisely in comparison to the regularized Burnett approximation.
However, the approximation in a whole has not improved (see
Fig.~\ref{AnnPhysFig4}). {\it Thus, we can conclude that although
the partial summation method} (\ref{regprocedure101a}) {\it and}
(\ref{regprocedure101b}) {\it is capable of removing the Bobylev
instability, and reproducing qualitatively the exact
Chapman--Enskog solution in the short-wave domain, the exactness
does not increase monotonically with the depth of the
approximation} $N$. This drawback of the regularization procedure
indicates once again that an attempt to capture the lower-order
terms of the Chapman--Enskog procedure does not succeed in a
better approximation as a whole.

\subsection{The dynamic invariance} \label{DI_intro}

The starting points of all the approaches considered so far (exact
or approximate) is the Chapman--Enskog expansion. However, the
result of the summation does not involve the Knudsen number
$\epsilon$ explicitly and does not require the ``smallness" of
this parameter. Therefore, it makes sense to reformulate the
problem of the reduced description (for the Grad equations
(\ref{Grad101}) this amounts to the problem of constructing a
function $\sigma_k (u_k , p_k ,k)$) in a way where the parameter
$\epsilon$ does not appear at all. Further, in the framework of
such an approach, we can seek a method of explicit construction of
the function $\sigma_k (u_k , p_k ,k)$, which does not rely upon
the Taylor-like expansions as above.

In this section we introduce such an approach, considering again
the  illustrative example (\ref{Grad101}). These ideas will be
extensively used in the sequel, and they also constitute the basis
of the so-called method of invariant manifold for dissipative
systems \cite{GKTTSP94}.

Let us rewrite here (\ref{Grad101}) in the Fourier variables, and
cancel the parameter $\epsilon$:
\begin{eqnarray}
\label{Grad101a}
\partial_t p_k &=&-\frac{5}{3}ik u_k ,\\\nonumber
\partial_t u_k &=&-ik p_k -ik \sigma_k ,\\\nonumber
\partial_t \sigma_k &=&-\frac{4}{3}ik u_k
-\sigma_k .
\end{eqnarray}

The result of the reduction in the system (\ref{Grad101a}) amounts
to a function $\sigma_k (u_k , p_k ,k)$, which depends
parametrically on the hydrodynamic variables $u_k $ and $p_k $,
and also on the wave vector $k$. Due to the linearity of the
problem under consideration, this function depends linearly on
$u_k $ and $p_k $, and we can start with the form given by
(\ref{sigma101}):
\begin{equation}
\label{sigma101a}
\sigma_{k}(u_k , p_k ,k)
=ikAu_k -k^2 Bp_k ,
\end{equation}
where $A$ and $B$ are undetermined functions of $k$. Now, however,
we do not refer to a power series representation of these
functions as in (\ref{AB101}).

Given the form of the function $\sigma_{k}(u_k , p_k ,k)$
(\ref{sigma101a}), we can compute its time derivative in {\it two
} different ways. On one hand, substituting (\ref{sigma101a}) into
the right hand side of the third equation in the set
(\ref{Grad101a}), we derive:
\begin{equation}
\label{derivative101micro}
\partial_t^{\rm micro}\sigma_k =-ik\left(\frac{4}{3}+A\right)u_k +k^2 Bp_k .
\end{equation}
On the other hand, computing the time derivative and using the
first two equations (\ref{Grad101a}), we obtain:
\begin{eqnarray}
\label{derivative101macro}
\partial_t^{\rm macro} \sigma_k & =&\frac{\partial \sigma_k }{\partial u_k
}\partial_t u_k +\frac{\partial \sigma_k }{\partial p_k
}\partial_t p_k \\\nonumber
&=&ikA\left(-ikp_k -ik\sigma_k \right)-k^2 B\left(-\frac{5}{3}iku_k
\right)\\\nonumber
&=&ik\left(\frac{5}{3}k^2 B+k^2 A\right)u_k +k^2 \left(A-k^2 B\right)p_k .
\end{eqnarray}
Equating the expressions in the right hand sides of
(\ref{derivative101micro}) and (\ref{derivative101macro}), and
requiring that the resulting equality holds for any values of the
variables $u_k $ and $p_k $, we derive the following two algebraic
equations:
\begin{eqnarray}
\label{invariance101}
F(A,B,k)&=&-A-\frac{4}{3}-k^2 \left(\frac{5}{3}B+A^2
\right)=0,\\\nonumber
G(A,B,k)&=&-B+A\left(1-k^2 B\right)=0.
\end{eqnarray}
These are exactly the equations (\ref{system101}), which were
obtained after summation of the Chapman--Enskog expansion. Now,
however, we have reached the same result  without using the
expansion. Thus, (\ref{invariance101}) (or, equivalently,
(\ref{system101})) can be used as a  starting point for the
construction of the function (\ref{sigma101a}).

It is important to comment on the somewhat formal manipulations
which have led to (\ref{invariance101}). First of all, by the very
sense of the reduced description problem, we are looking for a set
of functions $\sigma_k $ which depend on time only through the
time dependence of the hydrodynamic variables $u_k $ and $p_k $.
That is, we are looking for a  set (\ref{sigma101a}), which is
parameterized with the values of the hydrodynamic variables.
Further, the two time derivatives, (\ref{derivative101micro}) and
(\ref{derivative101macro}), are relevant to the ``microscopic``
and the ``macroscopic" evolution within the set (\ref{sigma101a}),
respectively. Indeed, the expression in the right hand side of
(\ref{derivative101micro}) is just the value of the vector field
of the original Grad equations at the points of the set
(\ref{sigma101a}). On the other hand, (\ref{derivative101macro})
expresses the time derivative in terms of the reduced
(macroscopic) dynamics, which, in turn, is self-consistently
defined by the form (\ref{sigma101a}). Equations
(\ref{invariance101}) provide, therefore, the {\it dynamic
invariance condition of the reduced description} for the set
(\ref{sigma101a}): the function $\sigma_k (u_k (t) ,p_k (t) ,k)$
is a solution to both the full Grad system (\ref{Grad101a}) and to
the reduced system which consists of the first two (hydrodynamic)
equations. For this reason, equations (\ref{invariance101}) and
their analogs which will be obtained on similar reasoning, will be
called {\it the invariance equations}.

\subsection{The Newton method} \label{DI_N}

Let us concentrate on the problem of solving the invariance
equations (\ref{invariance101}). Clearly, if we are going to
expand the functions $A$ and $B$ into power series (\ref{AB101}),
we shall return to the Chapman--Enskog procedure. Now, however, we
see that the Chapman--Enskog expansion is just a method to solve
the invariance equations (\ref{invariance101}), and maybe not even
the optimal one.

Another possibility is to use {\it iterative}  methods. Indeed, we
shall apply  Newton's method. The algorithm is as follows: Let
$A_0 $ and $B_0 $ are  some initial approximations chosen for the
procedure. The correction, $A_1 =A_0 + \delta A_1 $ and $B_1 = B_0
+ \delta B_1 $, due to the Newton iteration is obtained upon a
linearization (\ref{invariance101}) around the approximation $A_0
$ and $B_0 $. Computing the derivatives, we can represent the
equation of the Newton iteration in matrix form:
\begin{eqnarray}
\label{1Nitertation101}
\left( \begin{array}{cc}
\frac{\partial F(A,B,k)}{\partial A}\left.\right|_{A=A_0 , B=B_0 }
 &
\frac{\partial F(A,B,k)}{\partial B}\left.\right|_{A=A_0 , B=B_0 }
\\
\frac{\partial G(A,B,k)}{\partial A}\left.\right|_{A=A_0 , B=B_0 }
&
\frac{\partial G(A,B,k)}{\partial B}\left.\right|_{A=A_0 , B=B_0 }
\\
\end{array} \right)
\left(\begin{array}{c}\delta A_1 \\ \delta B_1 \\\end{array}\right) \nonumber \\
+\left(\begin{array}{c}F(A_0 , B_0 ,k)\\G(A_0 , B_0 , k)\\\end{array}\right)=0.
\end{eqnarray}
where
\begin{eqnarray}
\label{Ngradient101}
\frac{\partial F(A,B,k)}{\partial A}\left.\right|_{A=A_0 , B=B_0
}&=&-\left(1+2k^2 A_0 \right),\\\nonumber
\frac{\partial F(A,B,k)}{\partial B}\left.\right|_{A=A_0 , B=B_0
}&=&-\frac{5}{3}k^2 ,\\\nonumber
\frac{\partial G(A,B,k)}{\partial A}\left.\right|_{A=A_0 , B=B_0
}&=&1-k^2 B_0 ,\\\nonumber
\frac{\partial G(A,B,k)}{\partial B}\left.\right|_{A=A_0 , B=B_0
}&=&-\left(1+k^2 A_0 \right).
\end{eqnarray}
Solving the system of linear algebraic equations,
we come to the first correction $\delta A_1 $ and $\delta B_1 $.
Further corrections are found iteratively:
\begin{eqnarray}
\label{N101}
A_{n+1}&=&A_n +\delta A_{n+1},\\\nonumber
B_{n+1}&=&B_n +\delta B_{n+1},
\end{eqnarray}
where $n\ge 0$, and
\begin{equation}
\label{Niteration101}
\left(\begin{array}{cc} -(1+2k^2 A_{n} )
 &
-\frac{5}{3}k^2
\\
1-k^2 B_{n}
&
-(1+k^2 A_{n} )
\\
\end{array} \right)
\left(\begin{array}{c}\delta A_{n+1} \\ \delta B_{n+1} \\\end{array}\right)
+\left(\begin{array}{c}F(A_n , B_n ,k)\\G(A_n , B_n ,
k)\\\end{array}\right)=0.
\end{equation}

Within the algorithm just presented, the problem is how to choose
the initial approximation $A_0 $ and $B_0 $. The recursion
(\ref{N101}) and (\ref{Niteration101}) is applicable formally to
any initial approximation. However, the convergence (if at all)
might be sensitive to the choice.

\begin{figure}[t]
\centering{
\includegraphics[width=100mm, height=80mm]{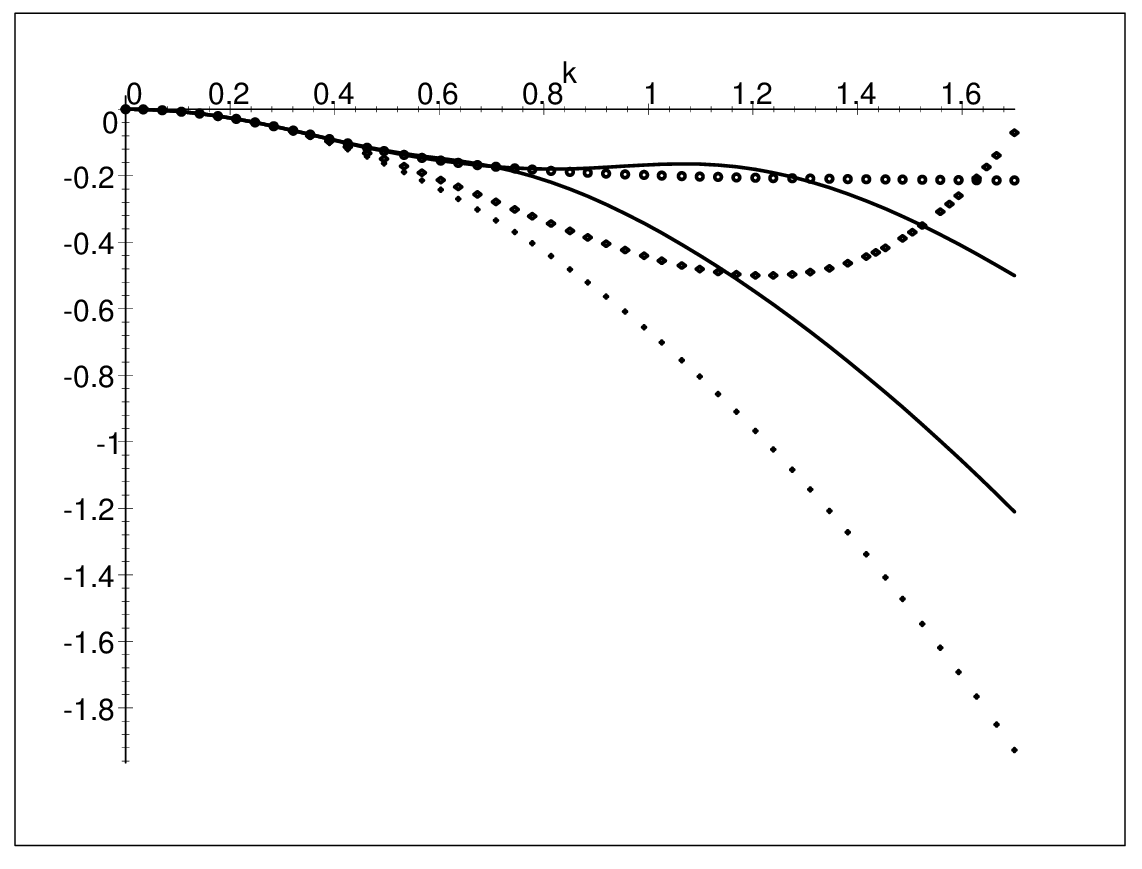}
\caption{\label{AnnPhysFig5}Attenuation rates for the Newton
method with the Navier--Stokes approximation as the initial
condition. Dots: The Navier--Stokes approximation. Solid: The
first and the second iterations of the invariance equation.
Circles: The exact solution to the invariance equation. Diamonds:
The super-Burnett approximation.}}
\end{figure}

For the first experiment let us take the Navier--Stokes
approximation of the functions $A$ and $B$:
\[ A_0 = B_0 =-\frac{4}{3}
\]
The outcome of the first two Newton iterations (the attenuation
rates as they follow from the first and second Newton iteration)
are presented in Fig.~\ref{AnnPhysFig5}. It is clearly seen that
the Newton iterations converge rapidly to the exact solution for
moderate $k$, but the asymptotic behavior in the short-wave domain
does not improve.

\begin{figure}[t]
\centering{
\includegraphics[width=100mm, height=80mm]{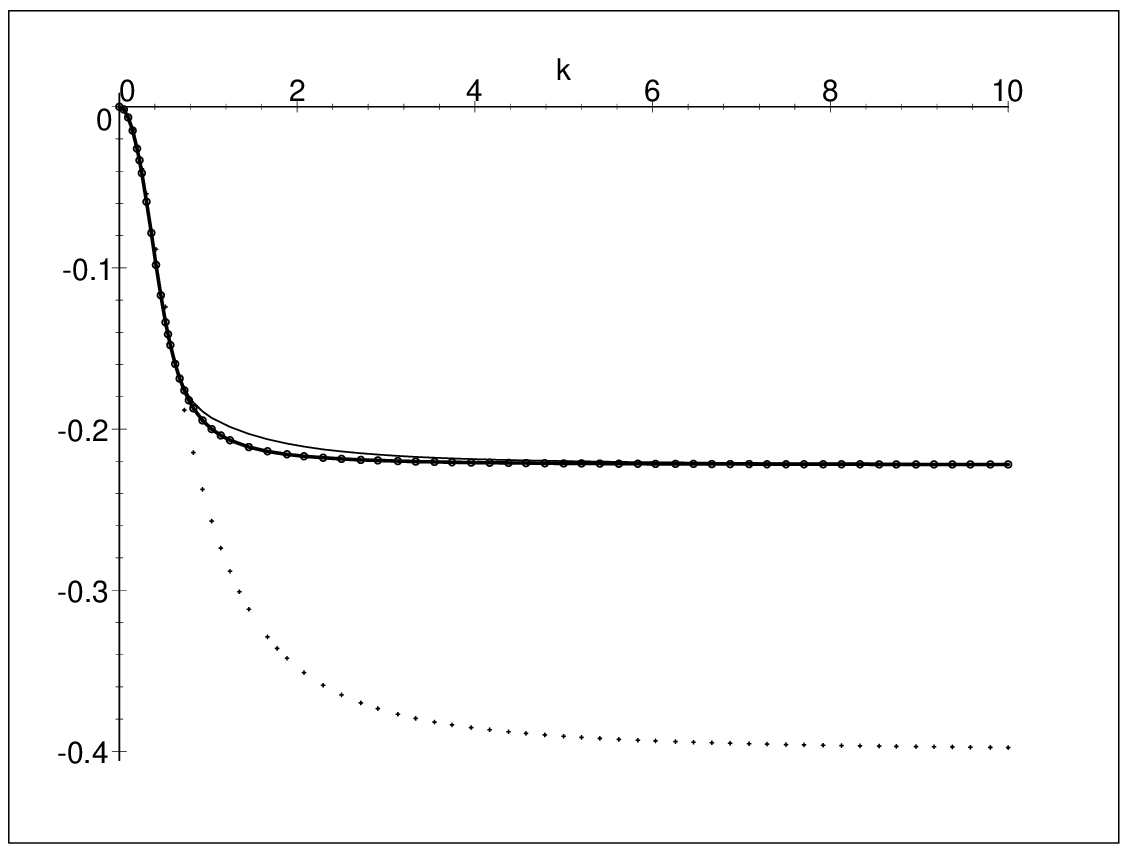}
\caption{\label{AnnPhysFig6}Attenuation rates with the regularized Burnett approximation as the
initial condition for the Newton method. Dots: The regularized Burnett approximation, or the first
Newton iteration with the Euler initial condition (see text). Solid: The first and the second Newton
iterations with the regularized Burnett approximation as the initial condition. Circles: The exact
solution to the invariance equation.}}
\end{figure}

Another possibility is to take the result of the regularization
procedure as presented above. Let the regularized Burnett
approximation (\ref{reg1AB101}) be taken for the initial
approximation, that is:
\begin{equation}
\label{initial101}
A_0 =A_1^R = -\frac{4}{3+5k^2 }, \quad B_0 = B_1^R
=-\frac{4}{3+5k^2 }.
\end{equation}
Substituting  (\ref{initial101}) into (\ref{N101}) and
(\ref{Niteration101}) for $n=0$ we obtain, after some algebra,
the following first correction:
\begin{eqnarray}
\label{1Nreg1101}
A_1 &=&-\frac{4(27+63k^2 +153k^2 k^2 +125k^2 k^2 k^2 )}{3(3+5k^2 )
(9+9k^2 +67k^2 k^2 +75 k^2 k^2 k^2 )},\\\nonumber
B_1 &=&-\frac{4(9+33k^2 +115k^2 k^2 +75 k^2 k^2 k^2 )}
{(3+5k^2 )(9+9k^2 +67k^2 k^2 +75 k^2 k^2 k^2 )}
\end{eqnarray}
Functions (\ref{1Nreg1101}) are not yet the exact solution to
(\ref{invariance101}) (that is, the functions $F(A_1, B_1,k)$ and
$G(A_1, B_1, k)$ are not equal to zero for all $k$). However,
substituting  $A_1 $ and $B_1 $ instead of $A$ and $B$ into the
dispersion relation (\ref{dispersion101}), we derive in the
short-wave limit:
\begin{equation}
\label{Nlimit101}
\lim_{|k|\rightarrow\infty}\omega_{\pm}=-\frac{2}{9}\pm
i|k|\sqrt{3}.
\end{equation}
That is, already the first Newton iteration, as applied to the
regularized Burnett approximation, leads to the exact expression
in the short-wave domain. Since the first Newton iteration appears
to be asymptotically exact, the next iterations improve the
solution only for the intermediate values of $k$, whereas the
asymptotic behaviour remains exact in all iterations. The
attenuation rates for the first and second Newton iterations with
the initial approximation (\ref{initial101}) are plotted in
Fig.~\ref{AnnPhysFig6}. The agreement with the exact solution is
excellent.

\begin{figure}[t]
\centering{
\includegraphics[width=100mm, height=80mm]{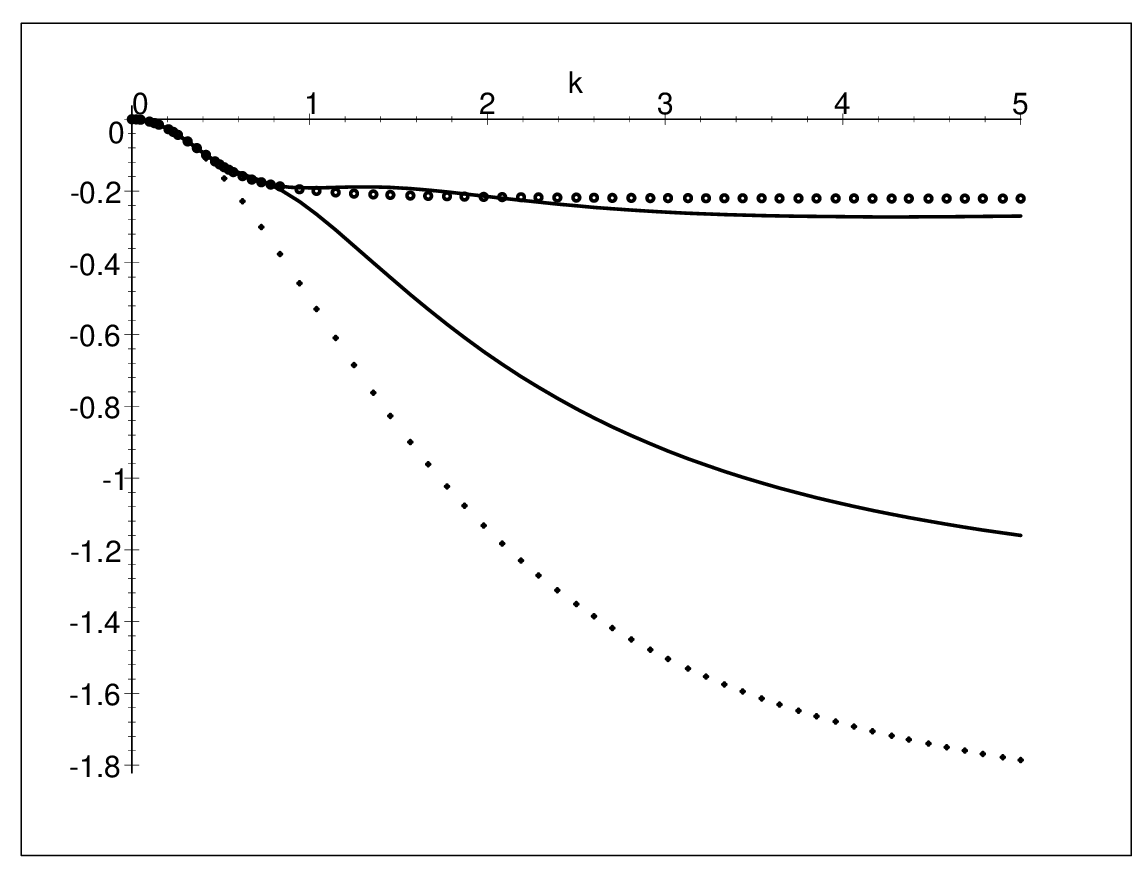}
\caption{\label{AnnPhysFig7}Attenuation rates with the regularized super-Burnett approximation as the
initial condition for the Newton method. Dots: The regularized super-Burnett approximation. Solid:
The first and the second Newton iterations. Circles: The exact solution to the invariance equation.}}
\end{figure}

One more test is to take the result of the super-Burnett
approximation (\ref{reg2AB101}) as an initial condition in the
Newton procedure (\ref{Niteration101}). As we know, the
regularization of the super-Burnett approximation provides a
poorer approximation in comparison to (\ref{initial101}),
particularly in the short-wave domain. Nevertheless, the Newton
iterations do converge though less rapidly (see
Fig.~\ref{AnnPhysFig7}).

The examples considered so far demonstrate that the Newton method,
as applied to the invariance equations (\ref{invariance101}) is a
more powerful tool in comparison to the Chapman--Enskog procedure.
It is also important that the initial approximation should be
``properly chosen", and that it should reproduce, at least
qualitatively, the features of the solution not only in the
long-wave limit, but over the whole range of wavenumbers.

The best from the initial approximations considered so far is the
regularized Burnett approximation (\ref{initial101}). We have
already commented on the relation of this approximation to the
invariance equations, as well as on its relation to the
Chapman--Enskog procedure. The further important observation  is
as follows:

Let us choose the {\it Euler} approximation for the functions $A$ and
$B$, that is:
\begin{equation}
\label{Euler101}
A_0 = B_0 =0
\end{equation}
The equation of the first Newton iteration (\ref{Niteration101}) is
very simple:
\begin{equation}\label{euleriteration}
\left(\begin{array}{cc} -1
 &
-\frac{5}{3}k^2
\\
1
&
-1
\\
\end{array} \right)
\left(\begin{array}{c}\delta A_{1} \\ \delta B_{1} \\\end{array}\right)
+\left(\begin{array}{c}-\frac{4}{3}\\0
\\\end{array}\right)=0,
\end{equation}
and
\begin{equation}
\label{N1E101}
A_1 =
B_1 = -\frac{4}{3+5k^2 }.
\end{equation}

Thus, {\it the regularized Burnett approximation is at the same
time the first Newton correction as applied to the Euler initial
approximation}. This property distinguishes the regularization of
the Burnett approximation from other regularizations. Now the
functions (\ref{1Nreg1101}) can be regarded as the  {\it second}
Newton correction as applied to the Euler initial approximation
(\ref{Euler101}).

Finally, let us examine what Newton's method does  in the case of
singularities. As we have demonstrated in the previous section,
the singularity of the diffusion-like mode occurs when this mode
couples to a non-hydrodynamic mode of the 10 moment Grad system if
the spatial dimension is greater that one.

\begin{figure}[t]
\centering{
\includegraphics[width=100mm, height=80mm]{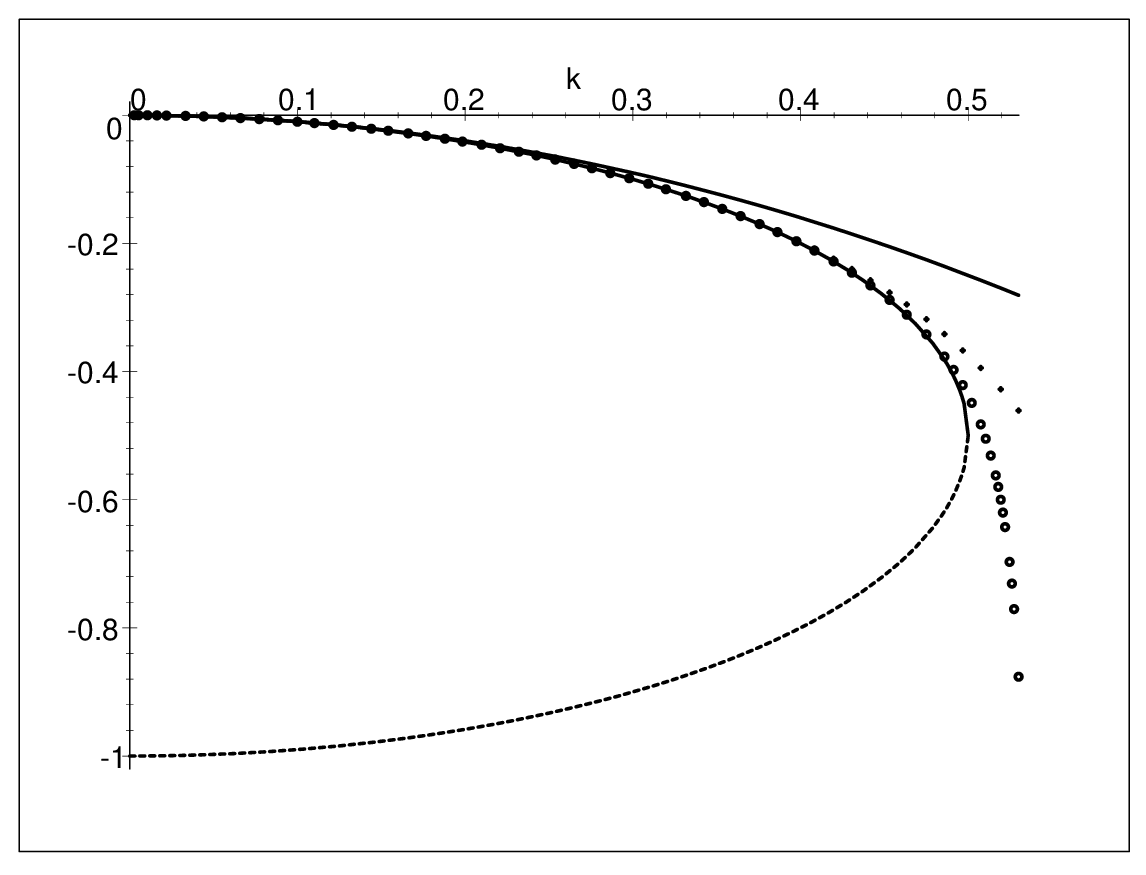}
\caption{\label{AnnPhysFig8}The diffusion mode with the Euler
initial approximation for the invariance equation. Upper solid:
The the first iteration. Dots: The second iteration. Circles: The
third iteration. Lower solid: The exact solution. Dash: The
critical mode. }}
\end{figure}

Without proving it here, the invariance equation method as applied
to the 10 moment Grad system (\ref{Grad103}) leads to the system
of equations (\ref{system103}). We have already demonstrated what
the outcome of the Newton method is when it is applied to the
first two equations of this system (responsible for the acoustic
mode and containing no singularities). The Newton method, as
applied to (\ref{quadratic103}), reads:
\begin{eqnarray}
\label{newton103}
Y_{n+1}&=&Y_n +\delta Y_{n+1},\\\nonumber
(1+2Y_n )\delta Y_{n+1}+\{Y_n (1+Y_n )+k^2 \}&=&0,
\end{eqnarray}
where $n\ge 0$, and $Y_0 $ is a chosen initial approximation. Taking
the Euler approximation ($Y_0 =0$), we derive:
\begin{eqnarray}
\label{Y12}
Y_1 &=&-k^2 ,\\\nonumber
Y_2 &=&-\frac{k^2 (1+ k^2 )}{1-2k^2 }.
\end{eqnarray}
The second approximation, $Y_2 $, is singular at $k_2
=\sqrt{1/2}$, and it can be demonstrated that all further
corrections also have the first singularity at points $k_n $, and
the sequence $k_2 ,\dots,k_n $ tends to the actual branching point
of the invariance equation (\ref{quadratic103}) $k_{\rm c} =1/2$.
The analysis of further corrections demonstrates that the
convergence is very rapid (see Fig.~\ref{AnnPhysFig8}).

The expressions (\ref{Y12}) demonstrate that unlike polynomial
approximations, Newton's method is capable of detecting the actual
singularities of the hydrodynamic spectrum. Formally, the function
$Y_2 $ becomes positive as $k$ becomes larger than $k_2 $, and
thus the attenuation rate, $\omega_d =Y_2 $ becomes positive after
this point. However, unlike the super-Burnett approximation for
the acoustic mode, this transition occurs now at a singular point.
Indeed, the attenuation rate $Y_2 $ tends to ``minus infinity", as
$k$ tends to $k_2 $ from the left. Thus, as described with the
Newton procedure, the non-physical domain is separated from the
physical one with an ``infinitely viscid" threshold. The
occurrence of the poles in the Newton iterations is, of course,
quite clear. Indeed, the Newton method involves the derivative of
the function $R(Y)=Y(Y+1)$ which appears on the left hand side of
(\ref{quadratic103}). The derivative $dR(Y)/dY$ becomes zero at
the singularity point $Y_c =-1/2$. The results of this section
bring us to the following conclusion:

(i) Exact summation of the Chapman--Enskog procedure results in
the same system of equations as the principle of dynamic
invariance. This was demonstrated above for a specific situation
but it holds for any (linearized) Grad system. The resulting
equations are always {\it nonlinear} (even for the simplest
linearized kinetic systems, such as Grad equations).

(ii) Now we are able to {\it alter the viewpoint}: the invariance
equations can be considered as basic in the theory, while the
Chapman--Enskog method is a way to solve it via an expansion in
powers of $k$. The method of power series expansion is neither the
only method to solve equations, nor the optimal. Alternative
iteration methods might be better suited to the problem of
constructing the reduced description.

(iii) An opportunity to derive the invariance equation in closed
form, and next to solve it this or that way is, of course, rather
exotic. The situation becomes complicated already for the
nonlinear Grad equations, and we should not expect anything simple
in the case of the Boltzmann equation. Therefore, if we are
willing to proceed along these lines in other problems, attention
should be drawn towards approximate procedures. With this, the
question arises: what amount of information is required to execute
the procedures? Indeed, the Navier--Stokes approximation can be
obtained without any knowledge of the whole nonlinear system of
invariance equations. It is important that the Newton method, as
applied to our problem, does not require any global information as
well. This was demonstrated above by a relation between the first
iteration as applied to the Euler approximation and the
regularization of the Burnett approximation.

\subsection{Invariance equation for the $1D13M$ Grad system} \label{DI_13_1}

Let us consider  as  the  next  example  the  problem  of   the   reduced
description for the one-dimensional thirteen moment Grad system.  Using
the dimensionless variables as above, we write the one-dimensional version
of the Grad  equations (\ref{balanceequations}) and (\ref{Grad133})
in the $k$-representation:
\begin{eqnarray}
\label{Grad131}
\partial_t \rho_k &=&-iku_k ,\\\nonumber
\partial_t u_k &=&-ik\rho_k -ikT_k -ik\sigma_k ,\\\nonumber
\partial_t T_k &=&-\frac{2}{3}iku_k -\frac{2}{3}ikq_k ,\\\nonumber
\partial_t   \sigma_k   &=&-\frac{4}{3}iku_k   -\frac{8}{15}ikq_k
-\sigma_k ,\\\nonumber
\partial_t q_k &=&-\frac{5}{2}ikT_k -ik\sigma_k -\frac{2}{3}q_k .
\end{eqnarray}
The Grad system (\ref{Grad131})  provides  the  simplest  coupling
of  the hydrodynamic  variables  $\rho_k  $,  $u_k   $,   and $T_k
$   to   the non-hydrodynamic variables, $\sigma_k $ and $q_k $,
the latter  corresponding to the heat flux. As above, our goal is
to reduce the description  of  the Grad system (\ref{Grad131}) to
the  three  hydrodynamic  equations  with respect to  the
variables  $\rho_k  $,  $u_k  $,  and  $T_k  $. That  is, we have
to express the functions $\sigma_k $ and $q_k $ in terms of $
\rho_k  $,  $u_k   $,   and   $T_k   $:
\begin{eqnarray*}
\sigma_k &=&\sigma_k (\rho_k ,u_k , T_k , k),\\
q_k &=&q_k (\rho_k ,u_k , T_k , k).
\end{eqnarray*}
Application of the Chapman--Enskog method in these cases, results
in the following algebraic scheme (we omit the Knudsen number
$\epsilon$):
\begin{eqnarray}
\label{procedure131}
\sigma_k^{(n)}&=&-\left\{\sum_{{m=0}}^{{n-1}}\partial_t^{(m)}\sigma_k^{(n-1-m
)}+\frac{8}{15}ikq_k^{(n-1)}\right\},\\ \nonumber
q_k^{(n)}&=&-\left\{\sum_{{m=0}}^{{n-1}}\partial_t^{(m)}q_k^{(n-1-m
)}+ik\sigma_k^{(n-1)}\right\},
\end{eqnarray}
where the Chapman--Enskog operators act as follows:
\begin{eqnarray}
\label{operators131}
\partial_t^{(m)}\rho_k &=&\left\{
\begin{array}{lr}-iku_k, &m=0\\0, &m\ge 1\end{array}\right.,\\\nonumber
\partial_t^{(m)}u_k &=&\left\{
\begin{array}{lr}-ik(\rho_k +T_k ),&m=0\\-ik\sigma_k^{(m-1)}, &m\ge
1\end{array}\right.,\\\nonumber
\partial_t^{(m)}T_k &=&\left\{
\begin{array}{lr}-\frac{2}{3}iku_k, &m=0\\-\frac{2}{3}ikq_k^{(m-1)}, &m\ge
1\end{array}\right..
\end{eqnarray}
The initial condition for the recurrence procedure
(\ref{procedure131}) reads:
$\sigma_k^{(0)}=-\frac{4}{3}iku_k          $,           and
$q_k^{(0)}=-\frac{15}{4}ikT_k$, which leads  to  the
Navier--Stokes-Fourier hydrodynamic equations.

Computing the coefficients $\sigma_k^{(1)}$ and $q_k^{(1)}$,
obtain the Burnett approximation:
\begin{eqnarray}
\label{burnett131}
\sigma_{1k} &=&-\frac{4}{3}iku_k +\frac{4}{3}k^2  \rho_k  -\frac{2}{3}k^2  T_k
,\\\nonumber
q_{1k}&=&-\frac{15}{4}ikT_k +\frac{7}{4}k^2 u_k .
\end{eqnarray}
The Burnett approximation (\ref{burnett131}) coincides with  that  obtained
from the Boltzmann equation,  and  it  is  precisely  the  case  where  the
instability was first demonstrated in the paper \cite{Bob}.

The structure of the terms  $\sigma_k^{(n)}$  and $q_k^{(n)}$ (an
analog  of  (\ref{structure101})  and  (\ref{structure103})) is as
follows:
\begin{eqnarray}
\label{structure131}
\sigma_k^{(2n)}&=&a_n (-k^2 )^n ik u_k ,\\\nonumber
\sigma_k^{(2n+1)}&=&b_n (-k^2 )^{n+1}\rho_n +c_n (-k^2  )^{n+1}T_k
,\\\nonumber
q_k^{(2n)}&=&\beta_n (-k^2)^n ik\rho_k + \gamma_n (-k^2)^n  ikiT_k
,\\\nonumber
q_k^{(2n+1)}&=&\alpha_n (-k^2 )^{n+1}  u_k .
\end{eqnarray}

The derivation of the invariance equation for the system
(\ref{Grad131}) goes along the same lines as in the previous
section. We seek the  functions  of the reduced description in the
form:
\begin{eqnarray}
\label{form131}
\sigma_k &=&ikAu_k -k^2 B \rho_k -k^2 CT_k ,\\\nonumber
q_k &=&ikX\rho_k +ikYT_k -k^2 Z u_k ,
\end{eqnarray}
where the functions $A,\dots, Z$ will be determinated in the
process.

The invariance condition results in a closed system of  equations  for  the
functions $A$, $B$, $C$, $X$, $Y$, and $Z$. As above, computing the microscopic time
 derivative
of the functions (\ref{form131}), due to the two last equations of the  Grad
system (\ref{Grad131}) we derive:
\begin{eqnarray}
\label{micro131}
\partial_t^{\rm micro}\sigma_k          &=&-ik\left(\frac{4}{3}-\frac{8}{15}k^2
Z+A\right)u_k \\\nonumber&&+k^2 \left(\frac{8}{15}X+B\right)\rho_k
+ k^2 \left(\frac{8}{15}Y+C\right)T_k ,\\ \nonumber
\partial_t^{\rm micro}q_k &=& k^2 \left(A+\frac{2}{3}Z\right)u_k +
ik\left(k^2   B-\frac{2}{3}X\right)\rho_k \\ \nonumber  -ik
\left(\frac{5}{2}-k^2   C -\frac{2}{3}Y\right)T_k .
\end{eqnarray}
On the other hand, computing the macroscopic time  derivative  due  to  the
first three equations of the system (\ref{Grad131}), we obtain:
\begin{eqnarray}
\label{macro131}
\partial_t^{\rm macro}\sigma_k       &=&\frac{\partial\sigma_k       }{\partial
u_k }\partial_t u_k                 +
\frac{\partial\sigma_k       }{\partial
\rho_k }\partial_t \rho +
\frac{\partial\sigma_k       }{\partial
T_k }\partial_t T_k \\\nonumber
&=&ik\left(k^2 A^2 +k^2 B+\frac{2}{3}k^2 C-\frac{2}{3}k^2 k^2  CZ\right)u_k
\\\nonumber
&&+\left(k^2 A-k^2 k^2 AB-\frac{2}{3}k^2 k^2 CX\right)\rho_k\\\nonumber
&&+\left(k^2 A-k^2 k^2 AC-\frac{2}{3}k^2 k^2 CY\right)T_k ;\\\nonumber
\partial_t^{\rm macro}q_k       &=&\frac{\partial q_k       }{\partial
u_k }\partial_t u_k                 +
\frac{\partial q_k       }{\partial
\rho_k }\partial_t \rho u_k +
\frac{\partial q_k       }{\partial
T_k }\partial_t T_k \\\nonumber
&=&\left(-k^2 k^2 ZA+k^2 X+\frac{2}{3}k^2 Y-\frac{2}{3}k^2 k^2 YZ\right)u_k
\\\nonumber
&&+ik\left(k^2 Z-k^2 k^2 ZB +\frac{2}{3}k^2 YX\right)\rho_k\\\nonumber
&&+ik\left(k^2 Z-k^2 k^2 ZC+\frac{2}{3}k^2 Y^2 \right)T_k .\\\nonumber
\end{eqnarray}
Equating the corresponding expressions in the formulas
(\ref{micro131}) and (\ref{macro131}), we derive the following
system of coupled equations:
\begin{eqnarray}
\label{invariance131} \nonumber F_1 &=&-\frac{4}{3}+\frac{8}{15}k^2  Z-A  -k^2  A^2  -k^2
B-\frac{2}{3}k^2 C+\frac{2}{3}k^2 k^2 CZ=0,\\\nonumber F_2 &=&\frac{8}{15}X+B-A+k^2 AB+\frac{2}{3}k^2
CX =0,
\\\nonumber F_3 &=&\frac{8}{15}Y+C-A+k^2 AC+\frac{2}{3}k^2 CY =0,  \\\nonumber F_4
&=&A+\frac{2}{3}Z+k^2 ZA-X-\frac{2}{3}Y+\frac{2}{3}k^2 YZ=0,\\\nonumber F_5 &=&k^2 B-\frac{2}{3}X-k^2
Z+k^2 k^2 ZB-\frac{2}{3}k^2 YX=0, \\ F_6 &=&-\frac{5}{2}+k^2 C -\frac{2}{3}Y-k^2 Z+k^2  k^2  ZC
-\frac{2}{3}k^2 Y^2 =0.
\end{eqnarray}
As above, the invariance equations  (\ref{invariance131})  can
also be obtained upon summation of the Chapman--Enskog expansion,
after the Chapman--Enskog procedure  is  casted  into a recurrence
relations   for   the coefficients $a_n ,\dots,\alpha_n $
(\ref{structure131}).  This route is less straightforward than the
one just presented, and  we omit the proof.

The Newton method, as applied to the system (\ref{invariance131}),
results in the following algorithm:

Denote  as  {\bfA}  the  six-component  vector  function   ${\bfA}
=(A,B,C,X,Y,Z)$. Let ${\bfA}_0 $ is the initial  approximation,
then:
\begin{equation}
\label{Newton131}
{\bfA}_{n+1}={\bfA}_n +\delta{\bfA}_{n+1},
\end{equation}
where $n\ge 0$, and the vector function $\delta{\bfA}_{n+1}$ is a
solution to the linear system of equations:
\begin{equation}
\label{iteration131}
{\bfN}_{n}\delta{\bfA}_{n+1}+{\bfF}_n =0.
\end{equation}
Here ${\bfF}_n $ is the vector function with the components  $F_i
({\bfA}_n )$, and ${\bfN}_n $ is a $6\times 6$ matrix:

\begin{eqnarray}
\label{gradient131}
\left(\begin{array}{cccc}
-(1+2k^2 A_n )  &  -k^2 &  -2/3k^2  (1-k^2  Z_n  )  & \\
k^2 B_n -1&1+k^2&2/3k^2 X_n & \\
k^2 C_n -1 &0&1+2/3k^2 Y_n +k^2 A_n & \\
1+k^2 Z_n &0&0& \\
0&k^2 (1+k^2 Z_n )&0& \\
0&0&k^2(1+k^2 Z_n )& \end{array}\right.  &&\\\nonumber
\left.\begin{array}{cccc} &0&0&2/3k^2  (4/5+k^2
C_n )\\
 &2/3(4/5+k^2 C_n )&0&0\\
 &0&2/3(4/5+k^2 C_n )&0\\
 &-1&-2/3(1-k^2 Z_n )&2/3+k^2 A_n +2/3k^2 Y_n \\
 &-2/3(1+k^2 Y_n )&-2/3k^2 X_n &-k^2 (1-k^2 B_n )\\
 &0&-2/3(1+2k^2 Y_n )&-k^2 (1-k^2 C_n )
\end{array}\right)&&
\end{eqnarray}

The Euler approximation gives: $A_0 =\dots =Z_0  =0$,  while  $F_1
=-4/3$, $F_6 =-5/2$, and $F_2 =\dots =F_5 =0$.  The  first  Newton
iteration  (\ref{iteration131})  as  applied   to   this   initial
approximation, leads  again to a  simple  algebraic  problem,  and
we have finally obtained:
\begin{eqnarray}
\label{1N131}
A_1&=&-20\frac{141k^2 +20}{867k^4 + 2105k^2 +300},\\\nonumber
B_1&=&-20\frac{459k^2 k^2 +810k^2 +100}{3468k^2 k^2 k^2
+12755k^2 k^2 +11725k^2 +1500},
\\\nonumber
C_1&=&-10\frac{51k^2 k^2 -485 k^2 -100}{3468k^2 k^2 k^2
+12755k^2 k^2 +11725k^2 +1500},
\\\nonumber
X_1&=&-\frac{375k^2 (21k^2 -5)}{2(3468k^2 k^2 k^2
+12755k^2 k^2 +11725k^2 +1500)},\\\nonumber
Y_1&=&-\frac{225(394k^2 k^2 +685 k^2 +100)}{4(
3468k^2 k^2 k^2 +12755k^2 k^2 +11725k^2 +1500)},\\\nonumber
Z_1&=&-15\frac{153k^2 +35}{867k^4 + 2105k^2 +300}.
\end{eqnarray}

Substituting (\ref{structure131}) into the first three equations
of the Grad system (\ref{Grad131}), and proceeding with the
dispersion relation as above, we derive the latter in terms of the
functions $A,\dots,Z$:
\begin{eqnarray}
\label{dispersion131}
\omega^3&-&k^2\left(\frac{2}{3} Y+ A\right) \omega^2 \\\nonumber
&+&k^2\left(\frac{5}{3}
-\frac{2}{3} k^2 Z-\frac{2}{3}k^2 C-k^2 B+\frac{2}{3}k^2 AY+\frac{2}{3} k^2 k^2 CZ
\right) \omega \\\nonumber &+&\frac{2}{3}k^2 (k^2 X-k^2 Y+k^2 k^2 BY -k^2 k^2 XC)=0.
\end{eqnarray}

When the functions $A_1 ,\dots,Z_1 $ (\ref{1N131}) are substituted
instead of $A,\dots,Z$ into  (\ref{dispersion131}), the dispersion
relation of the first Newton iteration, as applied to the
invariance equations (\ref{invariance131}) with the Euler initial
approximation, is obtained. This result coincides with the
regularization of the Burnett approximation, which was considered
in \cite{GKJETP91}. There it was demonstrate that the equilibrium
is stable within this approximation for arbitrary wave lengths.
The dispersion relation for the Burnett approximation, in turn, is
due to the approximation
\[ A=-4/3, \quad B=-4/3, \quad C=2/3,\quad X=0, \quad Y=-15/4, \quad Z=-7/4,\]
as it follows from a comparison of  (\ref{burnett131}) and
(\ref{form131}). The dispersion relation for the Burnett
approximation coincides with the one obtained in \cite{Bob} from
the Boltzmann equation.

\subsection{Invariance equation for the  $3D13M$ Grad system} \label{DI_13_3}

The final example to be considered is the 13 moment Grad system in
three spatial dimensions, (\ref{balanceequations}) and
(\ref{Grad133}). Let us rewrite here the original system in terms
of Fourier variables:
\begin{eqnarray}
\label{FGrad133}
\partial_t \rho\kkk &=&-ik\ek\cdot{\bfu}\kkk ,\\\nonumber
\partial_t {\bfu}\kkk &=&-ik\ek \rho\kkk -ik\ek T\kkk -ik \ek \cdot
\sk ,\\\nonumber
\partial_t T\kkk &=&-\frac{2}{3}ik(\ek\cdot{\bfu}\kkk+\ek\cdot{\bfq}
\kkk ),\\\nonumber
\partial_t \sk &=&-ik\overline{\ek{\bfu}\kkk}-\frac{2}{5}ik\overline{\ek {\bfq}\kkk}
-\sk
,\\\nonumber
\partial_t  {\bfq}\kkk &=&-\frac{5}{2}ik\ek T\kkk -ik\ek\cdot\sk -\frac{2}{3} {\bfq}\kkk .
\end{eqnarray}
Here we have represented the wave vector {\bfk} as ${\bfk}=k\ek$, and $\ek$ is the
unit vector.

The structure of the even and odd Chapman--Enskog coefficients,
$\s\kkk^{(n)}$ and $\bfq\kkk^{(n)}$, turns out to be as follows:
\begin{eqnarray}
\label{structure133} \nonumber \sk^{(2n)}&=&(-k^2 )^n ik\left\{a_n (\overline{\ek\uk}-2\gk
(\ek\cdot\uk ))+b_n \gk (\ek\cdot\uk )\right\},\\\nonumber \sk^{(2n+1)}&=&(-k^2 )^{n+1}\gk \{c_n
T\kkk +d_n \rho\kkk \},
\\\nonumber \bfq\kkk^{(2n)}&=&(-k^2 )^n ik\ek\{\gamma_n T\kkk +\delta_n \rho\kkk\},\\
\bfq\kkk^{(2n+1)}&=&(-k^2 )^{n+1}\{\alpha_n \ek (\ek\cdot\uk )+\beta_n (\uk -\ek (\ek\cdot\uk )\},
\end{eqnarray}
where $\gk =1/2\overline{\ek\ek}$, and the real-valued
coefficients $a_n ,\dots,\beta_n $ are due to the Chapman--Enskog
procedure (\ref{procedure133}) and (\ref{operators133}).

The expressions just presented suggest that the dynamic invariant
form of the stress tensor and of the heat flux reads:
\begin{eqnarray}
\label{form133} \sk&=&ikA (\overline{\ek\uk}-2\gk (\ek\cdot\uk ))+2ikB\gk (\ek\cdot\uk
)\\\nonumber&&- 2k^2 C\gk T\kkk -2k^2 D\gk\rho\kkk ,\\\nonumber \bfq\kkk&=&ikZ\ek T\kkk +ikU\ek
\rho\kkk\\\nonumber&& -k^2 X(\uk-\ek (\ek\cdot\uk ))-k^2 Y\ek (\ek\cdot\uk ),
\end{eqnarray}
where the functions $A,\dots ,Y$ depend on ${\bfk}$.
The dynamic invariance condition results in the following two closed systems for these
functions:
\begin{eqnarray}\label{invariance133-1}
\frac{2}{5}U+D-B+\frac{2}{3}k^2 CU +\frac{4}{3}k^2 BD&=&0,\\\nonumber
\frac{2}{5}Z+C-B+\frac{2}{3}k^2 CZ+\frac{4}{3}k^2 BC&=&0, \\\nonumber
-1+\frac{2}{5}k^2 Y-B-\frac{2}{3}k^2 C-k^2 D-\frac{4}{3}k^2 B^2 +\frac{2}{3}k^2
k^2 CY&=&0, \\\nonumber
\frac{4}{3}k^2 D-\frac{2}{3}U-k^2 Y-\frac{2}{3}k^2 ZU+\frac{4}{3}k^2 k^2 YD&=&0
,\\\nonumber
-\frac{5}{2}+\frac{4}{3}k^2 C-\frac{2}{3}Z-k^2 Y-\frac{2}{3}k^2 Z^2 +\frac{4}{3}
k^2 k^2 YC&=&0,\\\nonumber
\frac{4}{3} B +\frac{2}{3} Y- U-\frac{2}{3} Z+\frac{2}{3} k^2
ZY+\frac{4}{3} k^2 YB&=&0,
\end{eqnarray}
and
\begin{eqnarray}
\label{invariance133-2}
-1 -A +\frac{2}{5}k^2 X-k^2 A^2 &=&0,\\\nonumber
A+\frac{2}{3}X+k^2 AX&=&0
\end{eqnarray}
The method of summation of the Chapman--Enskog expansion can also
be developed, starting with the structure of the Chapman--Enskog
coefficients (\ref{structure133}), in the same manner as in
Sect.~\ref{exact_10}. Simple but rather extensive computations in
this case lead, of course, to the invariance equations
(\ref{invariance133-1}) and (\ref{invariance133-2}).

The Newton method, as applied to the systems
(\ref{invariance133-1}) and (\ref{invariance133-2}) with the
initial Euler approximation, leads in the first iteration to the
regularization of the Burnett approximation reported earlier in
\cite{GKJETP91}.

\begin{figure}[t]
\centering{
\includegraphics[width=100mm, height=80mm]{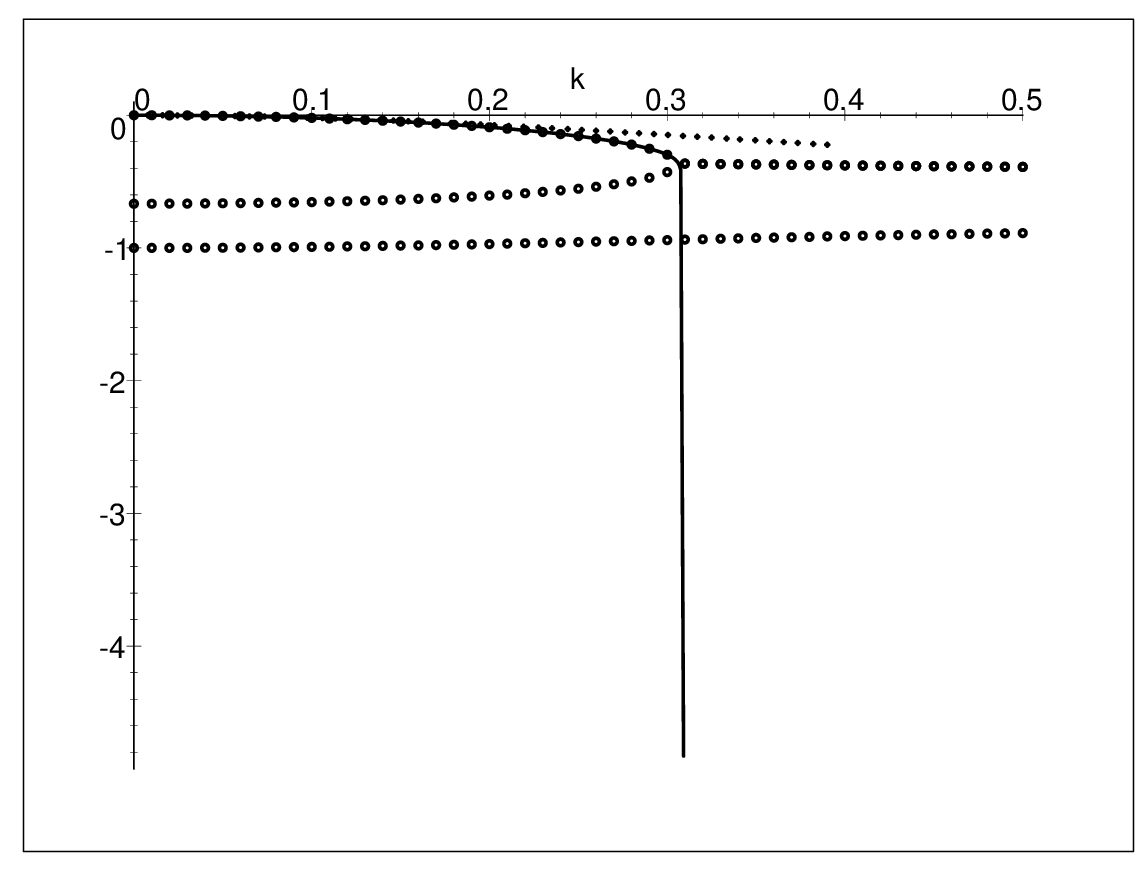}
\caption{\label{AnnPhysFig9}Solutions to  (\ref{invariance133-2}). Circles: Numerical solutions.
Dots: The first Newton iteration. Solid: The 4th Newton iteration.}}
\end{figure}

Introducing the functions $\bar{A}=k^2 A$  and $\bar{X}=k^2 X$ in
(\ref{invariance133-2}) we obtain:
\begin{equation}
\label{A133}
R(\bar{A})=\frac{5\bar{A}(3\bar{A}^2 +5\bar{A}+2)}{4(6\bar{A}+5)}=-k^2 ,
\end{equation}
while
\[\bar{X}=-\frac{3\bar{A}}{2+3\bar{A}}.\]
The derivative, $dR(\bar{A})/d\bar{A}$, becomes equal to zero for
$\bar{A}_c \approx -0.364$, which gives the critical wave vector
$k_{\rm c} =\sqrt{-R(\bar{A}_c )}\approx 0.305$. The Newton
method, as applied to  (\ref{A133}) with the initial Euler
condition $\bar{A}=0$, gives the following: the results of the
first and of the second iterations are regular functions, while
the third and the further iterations bring a singularity which
converges to the point $k_{\rm c} $ (see Fig.~\ref{AnnPhysFig9}).
These singularities (the real poles) of the Newton corrections are
of the same nature as discussed above.

\subsection{Gradient expansions in kinetic theory of phonons}
\label{DIP}

\subsubsection{Exact Chapman--Enskog solution and onset of second sound}

In this section, we close our discussion of linearized Grad
systems with an application to simple models for \index{Phonon
transport}phonon transport in rigid insulators. It is demonstrated
that the extended diffusion mode transforms into a second sound
mode due to its coupling to a non-hydrodynamic mode at some
critical value of the wave vector. This criticality shows up as a
branching point of the extension of the diffusion mode within the
Chapman--Enskog method. Although the analysis is essentially
similar to the examples considered above, it is presented in some
details for the sake of completeness.

Experiments on heat pulse propagation through crystalline media
\cite{experimenta,experimentb} confirmed the existence of a
temperature window (the Guyer-Krumhansl window
\cite{Krumhansl1,Krumhansl2,Krumhansl3}) with respect to which the
features of heat propagation are qualitatively different: At
temperatures exceeding the high-temperature edge of the window,
the heat propagates in a diffusion-like way. Below the
low-temperature edge of the window, the propagation goes in a
ballistic way, with the constant speed of sound. Within the
window, the propagation becomes wave-like. This latter regime is
called second sound (see \cite{Beck} for a review).

This  problem has drawn some renewed attention in the last years.
Models relevant for a unified description of diffusion, second
sound, and ballistic regimes of heat propagation are intensively
discussed (see \cite{Berlin,Ded} and references therein). To be
specific, recall the simplest and typical model of the phonon
transport \cite{Berlin}. Let $e({\bfx},t)$ and  ${\bfp}({\bfx},t)$
be small deviations of the energy density and energy flux of the
phonon field from their equilibrium values, respectively. Then

\begin{eqnarray}
\label{Grad}
\partial_t e &=&-c^2 \nabla\cdot {\bfp}, \label{equationa}\\
\partial_t {\bfp} &=&-\frac{1}{3}\nabla e -\frac{1}{\tau_R }{\bfp}.
\label{equationb}
\end{eqnarray}

Here $c$ is the \index{Debye velocity}Debye velocity of phonons,
and $\tau_R $ is the characteristic time of resistive processes.
Equations (\ref{Grad}) can be derived from the Boltzmann--Peierls
kinetic equation, within the relaxation time approximation,  by a
method similar to Grad's method \cite{Berlin}. Equations
(\ref{equationa}), (\ref{equationb}) provide the simplest model of
coupling between the hydrodynamic variable $e$ and the
non-hydrodynamic variable ${\bfp}$, allowing for a qualitative
description of both the diffusion and the second sound. Following
the standard argumentation \cite{Berlin}, we observe the two
limiting cases:
\begin{enumerate}
\item{As $\tau_R \to 0$, equation (\ref{equationb}) yields the
Fourier relation ${\bfp}=-\frac{1}{3}\tau_R \nabla e$ which closes
(\ref{equationa}) to give the diffusion equation:
\begin{equation}
\label{diffusion}
\partial_t e +\frac{1}{3}\tau_R c^2 \Delta e =0.
\end{equation}
} \item{As  $\tau_R \to \infty$, (\ref{equationb}) yields
$\partial_t {\bfp}=-\frac{1}{3}\nabla e$, and (\ref{equationa})
closes to give the wave equation:
\begin{equation}
\label{wave}
\partial_t^2 e +\frac{1}{3}c^2 \Delta e =0.
\end{equation}
} \end{enumerate}Equation (\ref{diffusion}) describes the usual
diffusive regime of heat propagation, while (\ref{wave}) is
relevant to the (undamped) second sound regime with the velocity
$u_2 =c/\sqrt{3}$; both are closed with respect to the variable
$e$.

However, even within the simplest model (\ref{equationa}),
(\ref{equationb}), the problem of closure remains unsolved in a
systematic way when $\tau_R $ is finite. The natural way of doing
so is provided by the Chapman--Enskog method. In the situation
under consideration, the Chapman--Enskog method yields an
extension of the diffusive transport to finite values of the
parameter $\tau_R $, and leads to an expansion of the
non-hydrodynamic variable {\bfp} in terms of the hydrodynamic
variable $e$. With this, if we are able to make this extension of
the diffusive mode exactly, we could learn more about the
transition between the diffusion and second sound (within the
framework of the model).

The Chapman--Enskog method, as applied to (\ref{equationa}),
(\ref{equationb}), results in the following series representation:
\begin{equation}
\label{series}
{\bfp}=\sum_{n=0}^{\infty}{\bfp}^{(n)},
\end{equation}
where the coefficients ${\bfp}^{(n)}$ are due to the
Chapman--Enskog recurrence procedure,

\begin{equation}
\label{procedure}
{\bfp}^{(n)}=-\tau_R \sum_{m=0}^{n-1}\partial_t^{(m)}{\bfp}^{(n-1-m)},
\end{equation}
while the Chapman--Enskog operators $\partial_t^{(m)}$ act on  $e$
as follows:
\begin{equation}
\label{operators}
\partial_t^{(m)}e=-c^2 \nabla\cdot {\bfp}^{(m)}.
\end{equation}
Finally, the zero order term reads:
${\bfp}^{(0)}=-\frac{1}{3}\tau_R \nabla e$, and leads to
the Fourier
approximation of the energy flux.

To sum up the series (\ref{series}) in closed form, we shall
specify the nonlinearity appearing in equations (\ref{procedure})
and (\ref{operators}). The coefficients ${\bfp}^{(n)}$ in
equations (\ref{series}) and (\ref{procedure}) have the following
explicit structure for arbitrary order $n\ge 0$:
\begin{equation}
\label{structure}
{\bfp}^{(n)}=a_n \Delta^n \nabla e,
\end{equation}
where the real-valued and yet unknown coefficients $a_n$ are due
to the recurrence procedure (\ref{procedure}), and
(\ref{operators}). Indeed, the form (\ref{structure}) is true for
$n=0$ ($a_0 =-\frac{1}{3}\tau_R $). Let us assume that
(\ref{structure}) is proven up to order $n-1$. Then, computing the
$n$th order coefficient ${\bfp}^{(n)}$, we derive:
\begin{eqnarray}
\label{induction}
{\bfp}^{(n)}&=&-\tau_R \sum_{m=0}^{n-1}\partial_t^{(m)}
a_{n-1-m}\Delta^{(n-1-m)}\nabla e\\\nonumber
&=&-\tau_R \sum_{m=0}^{n-1}a_{n-1-m}\Delta^{(n-1-m)}\nabla
\left(-c^2 a_m
\nabla\cdot\nabla \Delta^{m}e\right)\\\nonumber
&=&\tau_R c^2 \left\{\sum_{m=0}^{n-1}a_{n-1-m} a_m
\right\}\Delta^{n}\nabla e.
\end{eqnarray}

The last expression has the same form as (\ref{structure}). Thus,
the Chapman--Enskog procedure for the model (\ref{equationa}),
(\ref{equationb}) is equivalent to the following nonlinear
recurrence relation in terms of the coefficients $a_n $:

\begin{equation}
\label{recurrent}
a_n = \tau_R c^2 \sum_{m=0}^{n-1}a_{n-1-m} a_m ,
\end{equation}
subject to the initial condition  $a_0 =-\frac{1}{3}\tau_R $.
Further, it is convenient
to make the Fourier transform. Using ${\bfp}={\bfp}_{k}\exp\{i{\bfk}
\cdot{\bfx}\}$ and $e=e_{k}\exp\{i{\bfk}\cdot{\bfx}\}$, where
${\bfk}$ is the real-valued wave vector, we derive in (\ref{structure}):
${\bfp}^{(n)}_{k}=a_n i{\bfk}(-k^2 )^n e_{k},$
and
\begin{equation}
\label{AnnPhyssigma} {\bfp}_{k}=i{\bfk}A(k^2 )e_{k},
\end{equation}
where
\begin{equation}
\label{AB}
A(k^2 )=\sum_{n=0}^{\infty}a_n (-k^2 )^n .
\end{equation}
Thus, the the Chapman--Enskog solution (\ref{series}) amounts to
finding the function $A(k^2)$ represented by the power series
(\ref{AB}). If the function $A$ is known, the exact
Chapman--Enskog closure of the system (\ref{equationa}),
(\ref{equationb}) amounts to the following dispersion relation of
plane waves $\sim\exp\{\omega_{k}t+i{\bfk}\cdot {\bfx}\}$:
\begin{equation}
\label{dispersion}
\omega_{k}= c^2 k^2 A(k^2 ).
\end{equation}
Here $\omega_{k}$ is a complex-valued function of the real-valued
vector ${\bfk}$: ${\rm Re} (\omega_{k})$ is the attenuation rate,
${\rm Im} (\omega_{k})$ is the frequency.

Multiplying both equations in (\ref{recurrent}) with $(-k^2
)^{n}$, and performing a summation in $n$ from $1$ to infinity, we
get:
\[
A-a_0 =-\tau_R c^2 k^2 \sum_{n=0}^{\infty}\sum_{m=0}^{n}
a_{n-m}(-k^2 )^{n-m}a_m (-k^2 )^{m},
\]
Now we notice that
\[
\lim_{N\rightarrow\infty}\sum_{n=0}^{N}\sum_{m=0}^{n}
a_{n-m}(-k^2 )^{n-m}a_m (-k^2 )^{m}=A^2,
\]
Setting $a_0 = -\frac{1}{3}\tau_R $, we derive a quadratic
equation for the function $A$:
\begin{equation}
\label{system}
\tau_R  c^2 k^2 A^2 + A +\frac{1}{3}\tau_R =0.
\end{equation}
Further, a selection procedure is required to  choose the relevant
root of (\ref{system}). Firstly, recall that all the coefficients
$a_n $ (\ref{structure}) are real-valued by the sense of the
Chapman--Enskog method (\ref{procedure}) and (\ref{operators}),
hence the function $A$ (\ref{AB}) is real-valued. Therefore, only
the real-valued root of (\ref{system}) is relevant to the
Chapman--Enskog solution. The first observation is that
(\ref{system}) has no real-valued solutions as soon as $k$ becomes
bigger than the critical value $k_{\rm c} $, where
\begin{equation}
\label{critical}
k_{\rm c} =\frac{\sqrt{3}}{2\tau_R c}.
\end{equation}
Secondly, there are two real-valued solutions to (\ref{system}) at
$k<k_{\rm c} $. However, only one of them satisfies the
Chapman--Enskog asymptotic $ \lim_{k\to
0}A(k^2)=-\frac{1}{3}\tau_R . $

With these two remarks, we finally derive the following exact
Chapman--Enskog dispersion relation (\ref{dispersion}):
\begin{equation}
\label{root}
\omega_{k}=
\left\{ \begin{array}{ll}-(2\tau_R )^{-1}\left(1-\sqrt{1-
(k^2 )/(k_{\rm c}^2 ) }\right) & k <k_{\rm c} \\
{\rm none} & k > k_{\rm c} \end{array}\right. .
\end{equation}

This dispersion relation corresponds to the extended diffusion
transport, and it comes back to the standard Fourier approximation
in the limit of long waves  $k/k_{\rm c} \ll 1$. The
Chapman--Enskog solution does not exist as soon as $k/k_{\rm c}
>1$. For $k=k_{\rm c} $, the extended diffusion branch
crosses one of the  non-hydrodynamic branches of
(\ref{equationa}), (\ref{equationb}). For larger $k$, the extended
diffusion mode and the critical non-hydrodynamic mode produce a
pair of complex conjugate solutions with the real part equal to
$-\frac{1}{2\tau_R }$. The imaginary part of this extension after
$k_{\rm c}$ attains the asymptotic value $\pm iu_{2}k$, as $k\to
\infty$, where $u_2 =c/\sqrt{3}$ is the (undamped) second sound
velocity in the model (\ref{equationa}), (\ref{equationb}) (see
equation (\ref{wave})). Although the spectrum of the original
(\ref{equationa}), (\ref{equationb}) continues indeed after
$k_{\rm c} $, the Chapman--Enskog method does not recognize this
extension as part of the hydrodynamic branch, {\it while the
second sound regime is born from the extended diffusion after
coupling with the critical non-hydrodynamic mode.}

Finally, let us consider the opportunities provided by the Newton
method as applied to the invariance equation. First, the
invariance equation can be easily obtained in closed form here.
Consider again the expression for the heat flux in terms of the
energy density (\ref{AnnPhyssigma}), ${\bfp}_{k}=i{\bfk}A(k^2
)e_{k}$, where now the function $A$ is not thought as the
Chapman--Enskog series (\ref{AB}). The invariance equation is a
constraint on the function $A$, expressing the form-invariance of
the heat flux (\ref{AnnPhyssigma}) under both the dynamic
equations (\ref{equationa}) and (\ref{equationb}). Computing the
time derivative of function (\ref{AnnPhyssigma}) due to equation
(\ref{equationa}), we obtain:
\begin{equation}
\label{macro}
\partial_t^{\rm macro}{\bfp}_{k}=i{\bfk}A(k^2 )\partial_t e_{k}
=c^2 k^2 A^2 i{\bfk}e_{k}.
\end{equation}
On the other hand, computing the time derivative of the same function
due to equation (\ref{equationb}), we have:
\begin{equation}
\label{micro}
\partial_t^{\rm micro}{\bfp}_{k}=-\frac{1}{3}i{\bfk}e_{
k}-\frac{1}{\tau_R}Ai{\bfk}e_{k}.
\end{equation}
Equating (\ref{macro}) and (\ref{micro}), we derive the desired
invariance equation for the function A. This equation coincides
with the exact Chapman--Enskog equation (\ref{system}).

As the second step, let us apply the Newton method to the
invariance equation (\ref{system}), taking the Euler approximation
($A_0^{{N}} \equiv 0$) as the initial condition. Rewriting
(\ref{system}) in the form $F(A,k^2 )=0$, we come to the following
Newton iterations:
\begin{equation}
\label{Newton} \frac{\D F(A,k^2 )}{\D A}\bigg|_{A=A_n}(A_{n+1} -A_{n})+ F(A_n ,k^2 )=0.
\end{equation}
The first two iterations give:

\begin{eqnarray}
\label{iterations} \tau_R^{-1} A_1 &=&-\frac{1}{3}, \label{AnnPhysa}\\ \tau_R^{-1} A_2
&=&-\frac{1-\frac{1}{4}y^2 }{3(1-\frac{1}{2}y^2 )}. \label{AnnPhysb}
\end{eqnarray}

The first Newton iteration (\ref{AnnPhysa}) coincides with the
first term of the Chapman--Enskog expansion. The second Newton
iteration (\ref{AnnPhysb}) is a rational function with the Taylor
expansion coinciding with the Chapman--Enskog solution up to the
super-Burnett term, and it has a pole at $y_2=\sqrt{2}$. The
further Newton iterations are also rational functions with the
relevant poles at points $y_n$, and the sequence of this points
tends very rapidly to the location of the actual singularity $y_c
=1$ ($y_3\approx 1.17$, $y_4\approx 1.01$, etc.).

\subsubsection{Inclusion of  normal processes} \label{sec:normal}

Accounting for normal processes in the framework of the
semi-hydrodynamical models \cite{Berlin} leads to the following
generalization of (\ref{equationa}), (\ref{equationb}) (written in
Fourier variables for the one-dimensional case):
\begin{eqnarray}
\label{normal}
\partial_t e_{k}&=&-ikc^2 p_k , \label{normala}\\
\partial_t p_k&=&-\frac{1}{3}ike_k-ikN_k
-\frac{1}{\tau_R}p_k ,\label{normalb}\\
\partial_t N_k&=&-\frac{4}{15}ikc^2p_k-\frac{1}{\tau}N_k,\label{normalc}
\end{eqnarray}
where $\tau=\tau_N\tau_R/(\tau_N+\tau_R)$, $\tau_N$ is the
characteristic time of normal processes, and $N_k$ is the
additional field variable. Following the principle of invariance
as explained in the preceding section, we write the closure
relation for the non-hydrodynamic variables $p_k$ and $N_k$ as:
\begin{equation}
\label{closure2}
p_k=ikA_ke_k , \quad N_k=B_ke_k ,
\end{equation}
where $A_k$ and $B_k$ are two unknown functions of the wave vector
$k$. Further, following the principle of invariance as explained
in the previous section, each of the relations (\ref{closure2})
should be invariant under the dynamics due to  (\ref{normala}),
and due to (\ref{normalb}) and (\ref{normalc}). This  results in
two equations for the functions $A_k $ and $B_k $:
\begin{eqnarray}
\label{invariance20}
k^2c^2A_k^2&=&-\frac{1}{\tau_R}A_k-B_k-\frac{1}{3},\\\nonumber
k^2c^2A_kB_k&=&-\frac{1}{\tau}B_k+\frac{4}{15}k^2c^2A_k.
\end{eqnarray}
When the energy balance equation (\ref{normala}) is closed with the
relation (\ref{closure2}), this amounts to a dispersion relation for
the extended diffusion mode, $\omega_k=k^2c^2A_k$, where
$A_k$ is the solution to the invariance equations (\ref{invariance20}),
{\it subject to the condition} $A_k\to 0$ as $k\to 0$.
Resolving equations (\ref{invariance20}) with respect to $A_k$, and
introducing $\overline{A}_k=k^2c^2A_k$, we arrive at the following:
\begin{equation}
\label{invariance2}
\Phi(\overline{A}_k)=
\frac{5\overline{A}_k(1+\tau\overline{A}_k)(\tau_R \overline{A}_k+1)}
{5+9\tau\overline{A}_k}=-\frac{1}{3}\tau_R k^2c^2.
\end{equation}
The invariance equation (\ref{invariance2}) is completely analogous to the (\ref{system}). Written in
the form (\ref{invariance2}), it allows for a direct investigation of the critical points. For this
purpose,  we find  zeroes of the derivative, $d\Phi(\overline{A}_k)/d\overline{A}_k=0$. When the
roots of the latter equation, $\overline{A}_k^c$, are found, the critical values of the wave vector
are given as $-(1/3)k_{\rm c}^2c^2=\Phi(\overline{A}_k^c)$. The condition
$d\Phi(\overline{A}_k)/d\overline{A}_k=0$ reads:
\begin{equation}
\label{condition2}
18\tau^2\tau_R\overline{A}_k^3+3\tau(3\tau+8\tau_R)\overline{A}_k^2+
10(\tau+\tau_R)\overline{A}_k+5=0.
\end{equation}

Let us consider the particularly interesting case,
$\epsilon=\tau_N/\tau_R\ll 1$ (the normal events are less frequent
than resistive). Then the  real-valued root of (\ref{condition2}),
$\overline{A}_k(\epsilon)$, corresponds to the coupling of the
extended diffusion mode to the critical non-hydrodynamic mode. The
corresponding modification of the critical wave vector $k_{\rm c}$
(\ref{critical}) due to the normal processes amounts to  shifts
towards shorter waves, and we derive:
\begin{equation}
\label{critical2}
[k_{\rm c}(\epsilon)]^2=k_{\rm c}^2+\frac{3\epsilon}{10\tau_R^2c^2}.
\end{equation}

\subsubsection{Accounting for anisotropy} \label{sec:anisotropy}

The above examples concerned  the isotropic Debye model. Let us
consider the simplest anisotropic model of  a cubic media with a
longitudinal (L) and two degenerated transverse (T) phonon modes,
taking into account resistive processes only. Introduce the
Fourier variables, $e_{k}$, $e^{T}_{k}$, ${\bfp}_{k}^{T}$, and
${\bfp}_{k}^{L}$, where $e_{k}=e^{L}_{k}+2 e^{T}_{k}$ is the
Fourier transform of the total energy of the three phonon modes
(the only conserved quantity), while the rest of variables are
specific quantities. The isotropic model (\ref{equationa}),
(\ref{equationb}) generalizes to \cite{Berlin}:
\begin{eqnarray}
\label{anisotropy}
\partial_t e_{k}&=&-ic_L^2{\bfk}\cdot{\bfp}_{k}^L-
2ic_T^2{\bfk}\cdot{\bfp}_{k}^T,\label{anisotropy1}\\
\partial_t e_{k}^T&=&-ic_T^2{\bfk}\cdot{\bfp}_{k}^T+
\frac{1}{\lambda}\left[c_L^3(e_{k}-2e_{k}^{T})-c_{T}^3e_{
k}^T\right],\label{anisotropy2}\\
\partial_t {\bfp}_{k}^L&=&-\frac{1}{3}i{\bfk}(e_{k}-2e_{
k}^{T})-
\frac{1}{\tau_R^L}{\bfp}_{k}^L,\label{anisotropy3}\\
\partial_t {\bfp}_{k}^T&=&-\frac{1}{3}i{\bfk}e_{k}^{T}-
\frac{1}{\tau_R^T}{\bfp}_{k}^T,\label{anisotropy4}
\end{eqnarray}
where $\lambda=\tau_R^Tc_T^3+2\tau_R^Lc_L^3$.
The term containing
the factor $\lambda^{-1}$ corresponds to the energy exchange between
the L and T phonon modes.
The invariance constraint for the closure relations,

\begin{equation}
\label{closure3}
{\bfp}_{k}^L=i{\bfk}A_{k}e_{k},\quad
{\bfp}_{k}^T=i{\bfk}B_{k}e_{k},\quad
e_{k}^T=X_{k}e_{k},
\end{equation}
result in the following invariance equations for the {\bfk}-dependent
functions $A_{k}$, $B_{k}$, and $X_{k}$:

\begin{eqnarray}
\label{invariance3}
k^2c_L^2A_{k}^2+2k^2c_T^2A_{k}B_{k}&=&-\frac{1}{\tau_R^L}
A_{k}-\frac{1}{3}\left(1-2X_{k}\right),\label{inv31}\\
2k^2c_T^2B_{k}^2+k^2c_LB_{k}A_{
k}&=&-\frac{1}{\tau_R^T}B_{k}-\frac{1}{3}X_{k},
\label{inv32}\\
X_{k}\left(k^2c_L^2A_{k}+2k^2c_T^2B_{k}\right)&=&
c_T^2k^2B_{k}+\frac{1}{\lambda}\left[c_L^3-X_{k}
\left(2c_L^3+c_T^3\right)
\right].
\end{eqnarray}
When the energy balance equation (\ref{anisotropy1}) is closed
with the relations (\ref{closure3}), we obtain the dispersion
relation for the extended diffusion mode,
$\omega_{k}=\overline{A}_{k}+2\overline{B}_{k}$, where the
functions $\overline{A}_{k}=k^2c_L^2A_{k}$, and $\overline{B}_{k}
=k^2c_T^2B_{k}$, satisfy the conditions: $\overline{A}_{k}\to 0$,
and $\overline{B}_{k}\to 0$, as $k\to 0$. The resulting dispersion
relation is rather complicated in the general case of the four
parameters of the problem, $c_L$, $c_T$, $\tau_R^L$ and
$\tau_R^T$. Therefore, introducing the function $\overline{Y}_{
k}=\overline{A}_{k}+2\overline{B}_{k}$, let us consider the
following specific situations of closed equations for the
$\overline{Y}_{ k}$ on the basis of the invariance equations
(\ref{invariance3}):

(i) $c_L=c_T=c$, $\tau_R^L=\tau_R^T=\tau_R$ (complete degeneration
of the parameters of the L and T subsystems): The system
(\ref{invariance3}) results in two decoupled equations:
\begin{eqnarray}
\label{dispersioni}
\overline{Y}_{k}\left(\tau_R\overline{Y}_{
k}+1\right)+\frac{1}{3}k^2c^2\tau_R&=&0,\label{dispersioni1}\\
\left(
\tau_R\overline{Y}_{k}+1\right)^2+\frac{1}{3}k^2c^2\tau_R^2
&=&0.\label{dispersioni2}
\end{eqnarray}
Equation (\ref{dispersioni1}) coincides with (\ref{system}) for
the isotropic case, and its solution defines the coupling of the
extended diffusion to a non-hydrodynamic mode. Equation
(\ref{dispersioni2}) does not have a solution with the required
asymptotic behavior $\overline{Y}_{ k}\to 0$ as $k\to 0$, and is
therefore irrelevant to the features of the diffusion mode in this
completely degenerated case. It describes the two further
propagating and damped non-hydrodynamic modes of the
 (\ref{anisotropy}). The nature of these modes, as well of the mode which couples to the
diffusion mode well be seen below.

(ii) $c_L=c_T=c$, $\tau_R^L\ne\tau_R^T$ (nondegenerate
characteristic time of resistive processes in the $L$ and  $T$
subsystems):
\begin{eqnarray}
\label{dispersionii} && \left[\overline{Y}_{k}
\left(\tau_R^L\overline{Y}_{k}+1\right)+\frac{1}{3}k^2c^2\tau_R^L\right] \times
\left[\left(\tau_R^{\prime}\overline{Y}_{k}+3\right) \left(\tau_R^T\overline{Y}_{
k}+1\right)+\frac{1}{3}k^2c^2\tau_R^T\tau_R^{\prime}\right] \nonumber \\
&&+\frac{2}{3}k^2c^2\left(\tau_R^T-\tau_R^L\right)=0,
\end{eqnarray}
where $\tau_R^{\prime}=2\tau_R^L+\tau_R^T$. As
$\tau_R^T-\tau_R^L\to 0$, (\ref{dispersionii}) tends to the
degenerated case (\ref{dispersioni}). At $k=0$,
$\tau_R^L\ne\tau_R^L$, there are four solutions to
(\ref{dispersionii}). The $\overline{Y}_{0}=0$ is the hydrodynamic
solution indicating the beginning of the diffusion mode. The two
non-hydrodynamic solutions, $\overline{Y}_{0}=-1/\tau_R^L$, and
$\overline{Y}_{0}=-1/\tau_R^T$,
$\overline{Y}_{0}=-3/\tau_R^{\prime}$, are associated with the
longitudinal and  the transverse phonons, respectively. The
difference in relaxational times makes the latter transverse root
nondegenerate, and the third non-hydrodynamic mode,
$\overline{Y}_{0}=-3/\tau_R^{\prime}$, appears instead.

(iii) $c_L\ne c_T$, $\tau_R^L=\tau_R^T=\tau_R$ (nondegenerate
speed of the L and the T sound).
\begin{eqnarray}
\label{dispersioniii} &&\left[\overline{Y}_{k}\left(\tau_R\overline{Y}_{
k}+1\right)+\frac{1}{3}k^2c_L^2\tau_R\right]\times\left[\left(
\tau_R\overline{Y}_{k}+1\right)^2+\frac{1}{3}k^2c_T^2\tau_R^2\right] \nonumber \\
&&+\frac{2}{3}k^2\tau_R\frac{c_L^3(c_T^2-c_L^2)}{2c_L^3+c_T^3}\left(\tau_R
\overline{Y}_{k}+1\right)=0.
\end{eqnarray}
As $c_T-c_L\to 0$, (\ref{dispersioniii}) tends to the degenerate
case (\ref{dispersioni}). However, this time the non-hydrodynamic
mode associated with the transverse phonons degenerates at $k=0$.

Thus, we are able to identify the modes in  (\ref{dispersioni1})
and (\ref{dispersioni2}). The non-hydrodynamic mode which couples
to the extended diffusion mode is associated with the longitudinal
phonons, and is the case (\ref{dispersioni1}). The case
(\ref{dispersioni2}) is due to the transverse phonons. In the
nondegenerate cases, (\ref{dispersionii}) and
(\ref{dispersioniii}), both pairs of modes become propagating
after a certain critical values of $k$, and the behavior of the
extended diffusion mode is influenced by all three
non-hydrodynamic modes just mentioned. It should be stressed,
however, that the second sound mode, which is the continuation of
the diffusion mode \cite{experimenta,experimentb}, is due to
(\ref{dispersioni1}). The results of the above analysis lead to
the following conclusion:

(i) The examples considered above indicate an interesting
mechanism of a {\it kinetic} formation of the second sound regime
from the extended diffusion with the participation of the
non-hydrodynamic mode. The onset of the propagating mode shows up
as the critical point of the extension of the hydrodynamic
solution into the domain of finite $k$, which was found within the
Chapman--Enskog and equivalent approaches. These results concern
the situation at the high-temperature edge of the Guyer--Krumhansl
window, and are complementary to the coupling between the
transversal ballistic mode and the second sound at the
low-temperature edge \cite{Ranninger}.

(ii) The crossover from the diffusion-like to the wave-like
propagation was previously found in \cite{P1a,P1b,P1c} in the
framework of exact Chapman--Enskog solution to the Boltzmann
equation for the Lorentz gas model \cite{Hauge}, and for similar
models of phonon scattering in anisotropic disordered media
\cite{P2}. The characteristic common feature of the models studied
in \cite{Hauge,P1a,P1b,P1c,P2} and the models \cite{Berlin} is the
existence of a  gap between the hydrodynamic (diffusive) and the
non-hydrodynamic components of the spectrum. Therefore, one can
expect that the destruction of the extended  diffusion is solely
due to the {\it existence} of this gap. In applications to the
phonon kinetic theory this amounts to the introduction of the
relaxation time approximation. In other words, we may expect that
the mechanism of crossover from  diffusion to second sound in the
simple models \cite{Berlin} is identical to what could be found
from the phonon--Boltzmann kinetic equation within the relaxation
time approximation. However, it should be noted that the original
(i.e., without the relaxation time approximation) phonon kinetic
equations are {\it gapless} (see, for example, \cite{Beck}). On
the other hand, most of the works on heat propagation in solids
{\it do} exploit the idea of the gap, since it is only possible to
speak of diffusion if such a gap exists. To conclude this point,
the following general hypothesis can be expressed: {\it the
existence of diffusion (and hence of the gap in the relaxational
spectrum) leads to its destruction through the coupling with a
non-hydrodynamic mode.}

\subsection{Nonlinear Grad equations} \label{NL}

In the preceding sections, the Chapman--Enskog and other methods
were probed explicitly for the linearized Grad equations far
beyond the usual Navier--Stokes approximation. This was possible,
first of all because the problem of the reduced description was
shaped into a rather simple {\it algebraic} form. Indeed, the
algebraic structure of the stress tensor
$\sk(\rho\kkk,\uk,T\kkk,{\bfk})$ and of the heat flux
$\qk(\rho\kkk,\uk,T\kkk ,{\bfk})$ was fairly simple. However, when
we attempt to extend the approach onto the nonlinear Grad
equations, the algebraic structure of the problem is no longer
simple. Indeed, when we proceed along the lines of the
Chapman--Enskog method, for example, the number of {\it types} of
terms, $\nabla{\bfu}$, $\nabla\nabla{\bfu}$, $(\nabla{\bfu})^2$,
$\nabla T\nabla\rho$, and so on, in the Chapman--Enskog
coefficients $\s^{(n)}$ and $\bfq^{(n)}$ demonstrates a
combinatorial growth with the order $n$.

Still, progress is possible if we impose some rules for the
selection of the relevant terms. As applied to the Chapman--Enskog
expansion, these selection rules prescribe that only contributions
arising from terms with a definite structure in each order
$\s^{(n)}$ and $\bfq^{(n)}$ should be retained, and all other
terms should be ignored. This approach can be linked again with
the partial summation rules for the perturbation series in
many-body theories, where usually terms with a definite structure
are summed instead of the whole series. Our viewpoint on the
problem of the extension of the hydrodynamics in the nonlinear
case can be expressed as follows: The exact extension seems to be
impossible, and, moreover, quite useless because of the lack of a
physical transparency. Instead, certain sub-series of the
Chapman--Enskog expansion, selected on clear physical grounds, may
lead to less complicated equations, which, at the same time,
provide an extension for a certain subclass of hydrodynamic
phenomena. This viewpoint is illustrated in this section by
considering a sub-series of the Chapman--Enskog expansion which
provides the dominating contribution when the flow velocity
becomes very large (and thus it is relevant to a high-speed
subclass of hydrodynamic phenomena such as strong shock waves).

The approach to the Chapman--Enskog series for the nonlinear Grad
equations just mentioned, and which was based on a diagrammatic
representation of the Chapman--Enskog method, has been attempted
earlier in \cite{book}. In this section, however, we shall take
the route of the dynamic invariance equations which leads to the
same results more directly.

\subsubsection{The dynamic viscosity factor}

The starting point is the set of one-dimensional nonlinear
Grad equations
for the hydrodynamic variables
$\rho $, $u$ and $T$, coupled to the non-hydrodynamic variable  $\sigma$,
where $\sigma$ is the
$xx$-component of the stress tensor:

\begin{eqnarray}
\label{Gradnl}
\partial_t \rho &=&-\partial_x (\rho u); \label{Ga}\\
\partial_t u &=&-u\partial_x u -\rho^{-1}\partial_x p
-\rho^{-1}\partial_x \sigma; \label{Gb}\\
\partial_t T &=&-u\partial_x T-(2/3)T\partial_x u
-(2/3)\rho^{-1}\sigma\partial_x u; \label{Gc}\\
\partial_t \sigma &=&-u\partial_x \sigma -(4/3)p\partial_x u
-(7/3)\sigma\partial_x u -\frac{p}{\mu(T)}\sigma .\label{Gd}
\end{eqnarray}
Here $\mu(T)$ is the temperature-dependent viscosity coefficient.
We shall adopt the form $\mu(T)=\alpha T^{\gamma}$, which is
characteristic to the point-center models of particles collisions,
where $\gamma$ varies from $\gamma=1$ (the Maxwell molecules) to
$\gamma=1/2$ (hard spheres), and where $\alpha$ is a dimensional
factor.

Even in this model, the Chapman--Enskog expansion appears to be
exceedingly complicated in the full setting. Therefore,  we
address another, simpler problem: {\it What is the leading
correction to the Navier--Stokes approximation when the
characteristic value of the average velocity is comparable to the
thermal  velocity?}

Our goal is to compute the correction to the Navier--Stokes
closure relation, $\sigma_{\rm NS}=-(4/3)\mu\partial_x u$, for
high values of the average velocity. Let us consider first the
Burnett correction from  (\ref{Gradnl}):
\begin{equation}
\label{Burnett}
\sigma_{\rm B} =-\frac{4}{3}\mu\partial_x u +\frac{8(2-\gamma)}{9}\mu^2
p^{-1}(\partial_x u)^2-\frac{4}{3}\mu^2 p^{-1}\partial_x
(\rho^{-1}\partial_x p).
\end{equation}
The correction of the desired type is given by the
\index{Nonlinear viscosity}nonlinear term proportional to
$(\partial_x u)^2$. Each further $n$th term of the Chapman--Enskog
expansion contributes, among others, a nonlinear term proportional
to $(\partial_x u)^{n+1}$. Such terms can be named {\it
high-speed} terms since they dominate the rest of the
contributions in each  order of the Chapman--Enskog expansion when
the characteristic average velocity is comparable to the heat
velocity. Indeed, if $U$ is a characteristic mean velocity, and
$u=U\overline{u}$, where $\overline{u}$ is dimensionless, then the
term $(\partial_x u)^{n+1}$ receives the factor $U^{n+1}$ which is
the highest possible order of $U$ among the terms available in the
$n$th order of the Chapman--Enskog expansion. Simple dimensional
analysis leads to the conclusion that such terms are of the form
$\mu g^{n}\partial_x u$, where $g=p^{-1}\mu\partial_x u$ is
dimensionless. Therefore, the Chapman--Enskog expansion for the
function $\sigma$ may be formally rewritten as:
\begin{equation}
\label{representation}
\sigma\!=\!-\!\mu\!\left\{\!\frac{4}{3}\!-\!\frac{8(2\!-\!\gamma)}{9}g\!
+\!
r_2g^2\!+\!
\dots\!+\!r_ng^n\!+\!\dots\!\right\}
\!\partial_xu\!+\!\dots\!
\end{equation}
The series in the brackets is the collection of the high-speed
contributions of interest, coming from {\it all} orders of the
Chapman--Enskog expansion, while the dots outside the brackets
stand for the terms of other nature. Thus, the high-speed
correction to the Navier--Stokes closure relation in the framework
of the Grad equations (\ref{Gradnl}) takes the form:
\begin{equation}
\label{R}
\sigma_{\rm nl}=-\mu R(g)\partial_x u ,
\end{equation}
where $R(g)$ is the function represented by a formal subsequence
of Chapman--Enskog terms in the expansion (\ref{representation}).
The function $R$ can be viewed also as a dynamic modification of
the viscosity $\mu$ due to the gradient of the average velocity.

We shall now turn to the problem of an explicit derivation of the
function $R$ (\ref{R}). Following the principle of dynamic
invariance, we first compute the microscopic derivative of the
function $\sigma_{\rm nl}$ by substituting (\ref{R}) into the
right hand side of (\ref{Gd}):
\begin{eqnarray}
\label{microNL}
\partial_t^{\rm micro}\sigma_{\rm nl}&=&
-u\partial_x \sigma_{\rm nl}-\frac{4}{3}p\partial_x u -\frac{7}{3}\sigma_{\rm nl}\partial_x u
-\frac{p}{\mu(T)}\sigma_{\rm nl} \nonumber \\
&=&\left\{-\frac{4}{3}+\frac{7}{3}gR+R\right\}p\partial_x u +\dots,
\end{eqnarray}
where dots denote the terms irrelevant to the closure relation
(\ref{R}) (such terms appear, because (\ref{R}) is not the exact
closure relation).

Second, computing the macroscopic derivative of the closure
relation (\ref{R}) due to (\ref{Ga}), (\ref{Gb}), and (\ref{Gc}),
we obtain:
\begin{equation}
\label{macroNL1}
\partial_t^{\rm macro}\sigma_{\rm nl}=-[ \partial_t \mu(T)]R\partial_x
u-\mu(T)\frac{\D R}{\D g}[\partial_t g]\partial_x u- \mu(T)R\partial_x [\partial_t u].
\end{equation}
In the latter expression, the time derivatives of the hydrodynamic
variables should be replaced with the right hand sides of
(\ref{Ga}), (\ref{Gb}), and (\ref{Gc}), where, in turn, the
function $\sigma$ should be replaced by the function $\sigma_{\rm
nl}$ (\ref{R}). After some computation, we derive the following:

\begin{equation}
\label{macroNL2}
\partial_t^{\rm macro}\sigma_{\rm nl}=\left\{gR+\frac{2}{3}(1-gR)\times
\left(\gamma gR+(\gamma-1)g^2 \frac{\D R}{\D g}\right)\right\}p\partial_x u +\dots
\end{equation}
Again, the dots stand  for the terms irrelevant to the present
analysis.

Equating the relevant terms in (\ref{microNL}) and
(\ref{macroNL2}), we obtain the invariance equation for the
function $R$:
\begin{equation}
\label{AnnPhysinvariance} (1-\gamma)g^2 \left(1-gR\right)\frac{\D R}{\D g}+\gamma g^2 R^2
+\left[\frac{3}{2}+g(2-\gamma)\right]R-2=0.
\end{equation}

\begin{figure}[t]
\centering{
\includegraphics[width=100mm, height=80mm]{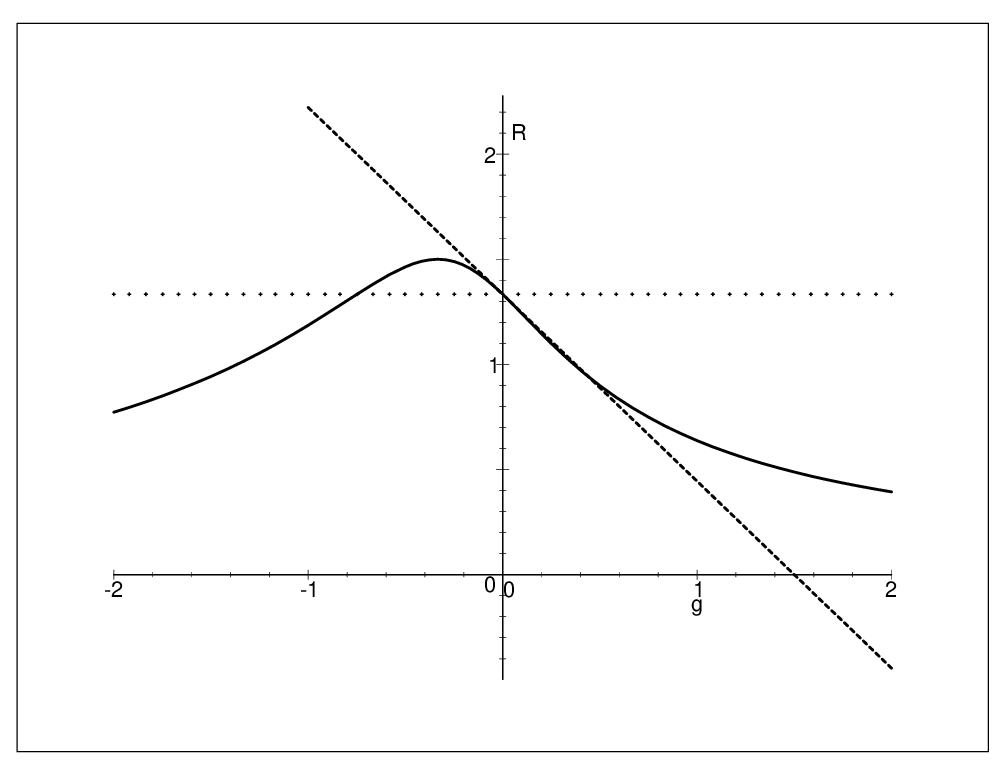}
\caption{\label{AnnPhysFig10}Viscosity factor $R(g)$ for Maxwell
molecules. Solid: exact solution. Dash: the Burnett approximation.
Dots: the Navier--Stokes approximation.}}
\end{figure}

For Maxwell molecules ($\gamma=1$), (\ref{AnnPhysinvariance})
simplifies considerably, and becomes algebraic:
\begin{equation}
\label{Maxwell}
g^2 R^2 +\left(\frac{3}{2}+g\right)R-2=0.
\end{equation}
The solution which recovers the Navier--Stokes closure relation in
the limit of small $g$ then reads:
\begin{equation}
\label{viscosity}
R_{\rm MM}=\frac{-3-2g+3\sqrt{1+(4/3)g+4g^2 }}{4g^2 }.
\end{equation}
Function $R_{\rm MM}$ (\ref{viscosity}) is plotted in
Fig.~\ref{AnnPhysFig10}. Note that $R_{\rm MM}$ is positive for
all values of its argument $g$, as is appropriate for the
viscosity factor, while the Burnett approximation to the function
$R_{\rm MM}$ violates positivity.

\begin{figure}[t]
\centering{
\includegraphics[width=100mm, height=80mm]{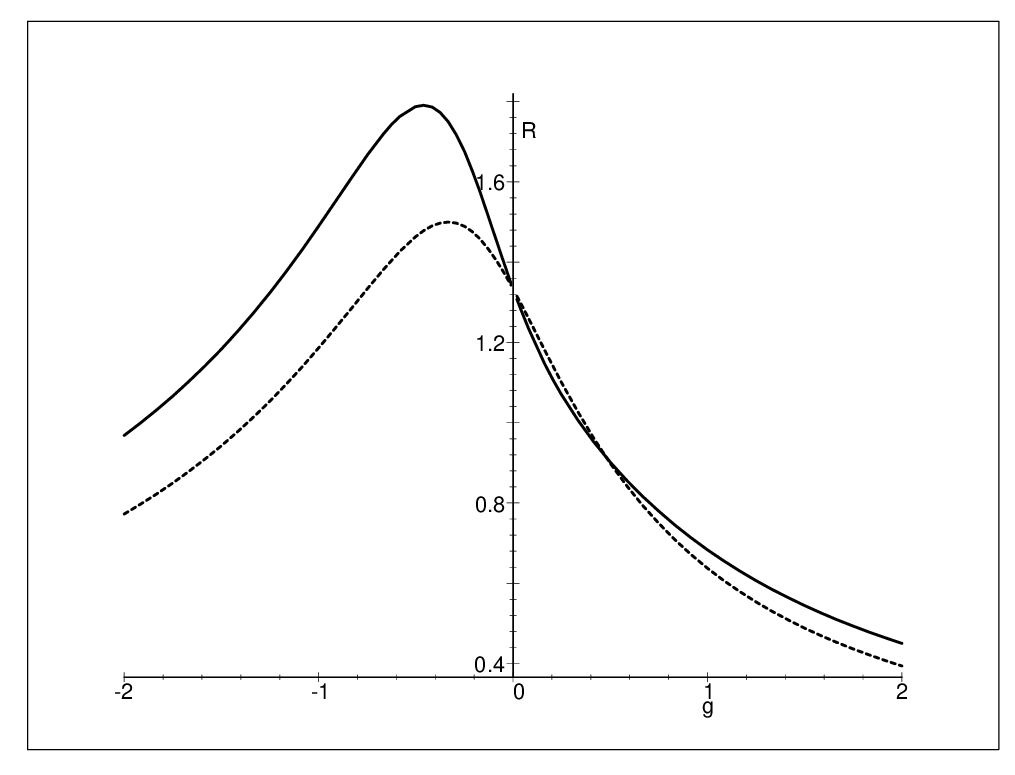}
\caption{\label{AnnPhysFig11}Viscosity factor $R(g)$ for hard spheres. Solid: the first
approximation. Dash: exact solution for the Maxwell molecules.}}
\end{figure}

For other models ($\gamma\ne 1$), the invariance equation
(\ref{AnnPhysinvariance}) is a rather complicated nonlinear ODE
with the initial condition $R(0)=4/3$ (the Navier--Stokes
condition). Several ways to derive analytic results are possible.
One possibility is to expand the function $R$ into powers of $g$,
in the point $g=0$. This will bring us back to the original
sub-series of the Chapman--Enskog expansion (see
(\ref{representation})). Instead, we take advantage of the
opportunity offered by the parameter $\gamma$. Introduce another
parameter $\beta=1-\gamma$, and consider the expansion:
\[ R(\beta,g)=R_0 (g)+\beta R_1 (g)+\beta^2 R_2 (g)+ \dots .\]
Substituting this expansion into the invariance equation
(\ref{AnnPhysinvariance}), we derive $R_0 (g)=R_{\rm MM}(g)$,
\begin{equation}
\label{approximation} R_1 (g)=-g(1-gR_0 )\frac{R_0 +g(\D R_0 /\D g)}{2g^2 R_0 +g +(3/2)},
\end{equation}
etc. That is, the  solution for other models is constructed in a
form of a series with the exact solution for the Maxwell molecules
as the leading term. For hard spheres ($\beta=1/2$), the result to
the first-order term reads: $R_{\rm HS}\approx R_{\rm
MM}+(1/2)R_1$. The resulting approximate viscosity factor is shown
in Fig.~\ref{AnnPhysFig11}. The features of the approximation
obtained are qualitatively the same as in the case of Maxwell
molecules.

It is interesting that precisely the same result for the nonlinear elongational viscosity obtained in
 \cite{KDNPRE97} was derived later by A. Santos \cite{Santos1} from the solution to the BGK
kinetic equation in the regime of so-called homoenergetic extension flow, see also
\cite{Santos2}, where a misprint in \cite{KDNPRE97} in formula (\ref{approximation}) was
detected. For further discussion, see \cite{Santos3}.

\subsubsection{Attraction to the invariant set}

Above, we have derived a correction to the Navier--Stokes
expression for the stress $\sigma$, in the one--dimensional case,
for large values of the average velocity $u$. This correction has
the form $\sigma=-\mu R(g)\partial_x u$, where $g\propto
\partial_x u$ is the longitudinal rate. The viscosity factor
$R(g)$ is a solution to the differential equation
(\ref{AnnPhysinvariance}), subject to a certain initial condition.
Uribe and Pi\~{n}a \cite{Uribe} have indicated some interesting
features of the invariance equation (\ref{AnnPhysinvariance}) for
the model of hard spheres. In particular, they have found that a
numerical integration from the initial point into the domain of
negative longitudinal rates is very difficult. What happens to the
relevant solution for negative values of $g$?

Let us denote as $P=(g,R)$ the points in the $(g,R)$ plane. The
relevant solution $R(g)$ emerges from the point $P_0=(0,4/3)$, and
can be uniquely continued to arbitrary values of $g$, positive and
negative. This solution can be constructed, for example, with the
Taylor expansion, and  which is identical with the relevant
sub--series of the Chapman--Enskog expansion. However, the
difficulty in constructing this solution numerically for $g<0$
originates from the fact that the same point $P_0$ is the point of
{\it essential singularity} of other (irrelevant) solutions to
(\ref{AnnPhysinvariance}). Indeed, for $|g|\ll 1$, let us consider
$\tilde{R}(g)=R(g)+\Delta$, where $R(g)=(4/3)+(8/9)(\gamma-2)g$ is
the relevant solution for small $|g|$, and $\Delta(g)$ is a
deviation. Neglecting in  (\ref{AnnPhysinvariance}) all regular
terms (of the order $g^2$), and also neglecting $g\Delta$ in
comparison to $\Delta$, we derive the following equation:
$(1-\gamma)g^2(\D \Delta/\D g)=-(3/2)\Delta$. The solution is
$\Delta(g)=\Delta(g_0)\exp[a(g^{-1}-g_0^{-1})]$, where
$a=(3/2)(1-\gamma)^{-1}$. The essential singularity at $g=0$ is
apparent from this solution, unless $\Delta(g_0)\ne 0$ (that is,
no singularity exists only for the relevant solution,
$\tilde{R}=R$). Let $\Delta(g_0)\ne 0$. If $g<0$, then $\Delta\to
0$, together with all its derivatives, as $g\to 0$. If $g>0$, the
solution blows up, as $g\to 0$.

The complete picture for $\gamma\ne1$ is as follows: The lines
$g=0$ and $P=(g,g^{-1})$ define the boundaries of the basin of attraction
$A=A_-\bigcup A_+$, where $A_-=\{P|-\infty<g<0,R>g^{-1}\}$, and
$A_+=\{P|\infty>g>0,R<g^{-1}\}$. The graph $G=(g,R(g))$ of the relevant
solution belongs to the closure of $A$, and goes through the points
$P_0=(0,4/3)$, $P_-=(-\infty,0)$, and $P_+=(\infty,0)$. These points
at the boundaries of $A$ are the points of essential singularity
of any other (irrelevant) solution with the initial condition
$P\in A$, $P\notin A\bigcap G$. Namely, if $P\in A_+$, $P\notin A_+\bigcap G$,
the solution blows up at $P_0$, and attracts to $P_+$. If
$P\in A_-$, $P\notin A_-\bigcap G$, the solution blows up at $P_-$, and
attracts to $P_0$.

\begin{figure}[t]
\centering{
\includegraphics[width=110mm, height=95mm]{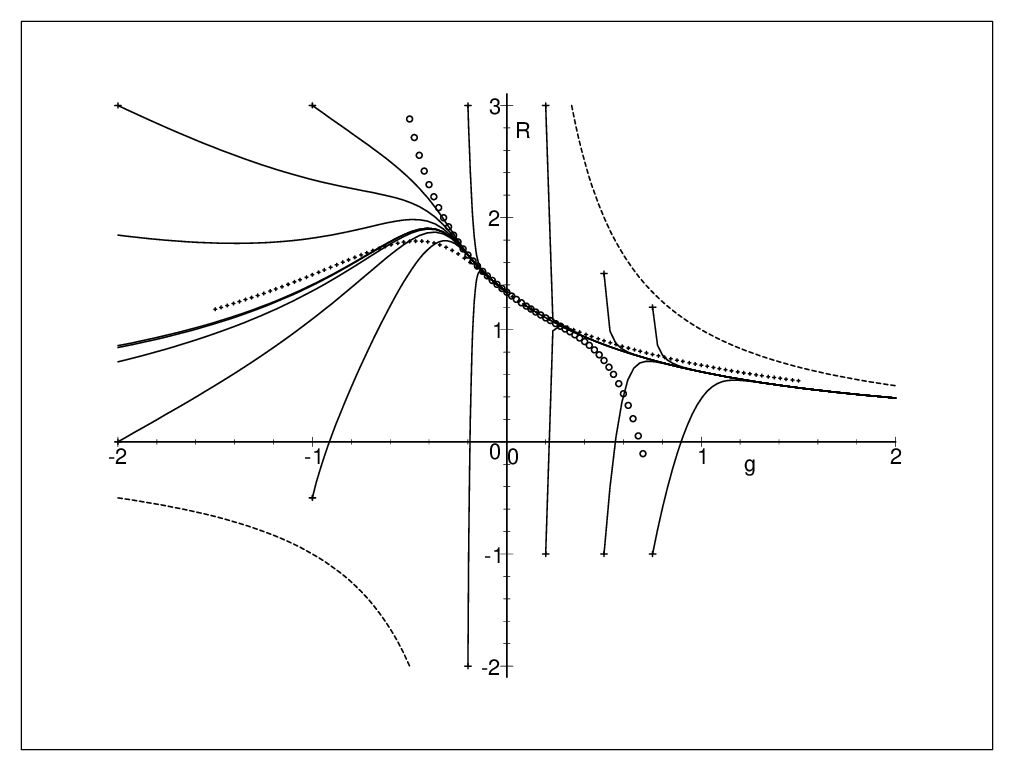}
\caption{\label{AnnPhysFig12}Viscosity factor as an attractor. Solid lines: numerical integration
with various initial points (crosses). Two poorly resolved lines correspond to the initial conditions
$(-100,0)$ and $(-100,3)$. Circles: Taylor expansion to the $5$th order. Dots: the analytical
approximation of (\ref{approximation}). Dash: boundaries of the basin of attraction.}}
\end{figure}

The above consideration is supported by a numerical study of
(\ref{AnnPhysinvariance}). In Fig.~\ref{AnnPhysFig12}, it is
demonstrated how the dynamic viscosity factor $R(g)$ emerges as
the attractor of various solutions to the invariance equation
(\ref{AnnPhysinvariance}) [the case considered corresponds to hard
spheres, $\gamma=1/2$]. The analytical approximation
(\ref{approximation}) is also shown in Fig.~\ref{AnnPhysFig12}. It
provides a reasonable global approximation to the attractor for
both positive and negative $g$. We conclude with a discussion.

(i) The main feature of the above  example of extending the
hydrodynamic description into a highly non-equilibrium and
nonlinear domain can be expressed as follows: this is an {\it
exact partial summation} of the Chapman--Enskog expansion.
``Partial" means that the relevant high-speed terms, dominating
the other contributions in the limit of the high average velocity,
were accounted to all orders of the original Chapman--Enskog
expansion. ``Exact" means that, though we have used the formally
different route,  the result is indeed  the sum of the relevant
sub-series of the original Chapman--Enskog expansion. In other
words, if we now expand the function $R_{\rm MM}(g)$
(\ref{viscosity}) in powers of $g$, around the point $g=0$, we
obtain again to the corresponding series inside the brackets in
(\ref{representation}). That this is indeed true can be checked up
to the few lower orders straightforwardly, although the complete
proof requires a more involved analysis. As the final comment to
this point, we would like to stress a certain similarity between
the problem considered above and the frequent situations in
many-body problems: there is no single leading {\it term};
instead, there is the leading {\it sub-series} of the perturbation
expansions, under certain conditions.

(ii) Let us discuss briefly the features of the resulting
hydrodynamics. The hydrodynamic equations are now given by
(\ref{Ga}), (\ref{Gb}), and (\ref{Gc}), where $\sigma$ is replaced
by $\sigma_{\rm nl}$ (\ref{R}). First, the correction concerns the
nonlinear regime, and, thus, the linearized form of the new
equations coincides with the linearized Navier--Stokes equations.
Second, the solution (\ref{viscosity}) for Maxwell molecules and
the result of the approximation (\ref{approximation}) for other
models suggest that the modified viscosity $\mu R$ gives a
vanishing contribution in the limit of very high values of the
average velocity. This feature seems to be of no surprise: if the
average velocity is very high in comparison to other
characteristic velocities (in our case, to the heat velocity), no
mechanisms of momentum transfer are relevant except for the
transfer with the stream. However, a cautious remark is in order
since the original ``kinetic" description are Grad's equations
(\ref{Gradnl}) and not the Boltzmann equation.

(iii) The invariance equation (\ref{AnnPhysinvariance}) defines
the relevant physical solution to the viscosity factor for all
values of $g$, and demonstrates an interesting phase-space
behavior similar to those of  finite-dimensional dynamical
systems.

\section{The main lesson} \label{CONCL}

Up to now, the problem of the exact relationship between kinetics
and hydrodynamics remains unsolved. All the methods used to
establish this relationship are not rigorous, and involve
approximations. In this work, we have considered situations where
hydrodynamics is the exact consequence of kinetics, and in that
respect, a new class of exactly solvable models of statistical
physics has been established.

The main lesson we can learn from the exact solution is the following: The
Chapman--Enskog method is the Taylor series expansion approach to solving the appropriate
invariance equation. Alternative iteration methods are much more robust for solving this
equation. Therefore, it seems quite important to develop approaches to the problem of
reduced description based on the principle of dynamic invariance \cite{GorKarBook2005}
rather than on particular methods of solving the invariance equations. The exact
solutions where these questions can be answered in all the quantitative details provide a
sound motivation for such developments.

\end{document}